\documentclass[fleqn,10pt]{wlscirep}
\usepackage[utf8]{inputenc}
\usepackage[T1]{fontenc}
\usepackage[normalem]{ulem}
\usepackage{multirow}
\usepackage{lineno}
\newcommand{\blue}[1]{{\color{blue}#1}}

\title{
Polaron-enhanced polariton nonlinearity \\
in lead halide perovskites \\
}

\author[1,$^{\dagger}$]{Mikhail~A.~Masharin}
\author[1,$^{\dagger}$]{Vanik~A.~Shahnazaryan}
\author[1]{Fedor~A.~Benimetskiy}
\author[1,2]{D.~N.~Krizhanovskii}
\author[1,3]{Ivan~A.~Shelykh}
\author[1]{Ivan~V.~Iorsh}
\author[1]{Sergey~V.~Makarov}
\author[1,*]{Anton~K.~Samusev}

\affil[1]{ITMO University, School of Physics and Engineering, St. Petersburg, 197101, Russia}
\affil[2]{Department of Physics and Astronomy, University of Sheffield, Sheffield S3 7RH, United Kingdom}
\affil[3]{Science Institute, University of Iceland, Dunhagi 3, IS-107, Reykjavik, Iceland}
\affil[*]{Corresponding author: anton.samusev@gmail.com}
\affil[$^{\dagger}$]{These authors contributed equally: Mikhail A.~Masharin, Vanik A.~Shahnazaryan}

\begin{abstract}
Exciton-polaritons offer a versatile platform for realization of all-optical integrated logic gates due to the strong effective optical nonlinearity resulting from the exciton-exciton interactions.
In most of the current excitonic materials there exists a direct connection between the exciton robustness to thermal fluctuations and the strength of exciton-exciton interaction, making materials with highest levels of exciton nonlinearity applicable at cryogenic temperatures only. 
Here, we show that strong polaronic effects, characteristic for perovskite materials, allow to overcome this limitation.
Namely, we demonstrate the record-high value of the nonlinear optical response in nanostructured organic-inorganic halide perovskite MAPbI$_3$, experimentally detected as 19.7~meV blueshift of the polariton branch under femtosecond laser irradiation.
This is substantially higher than characteristic values for the samples based on conventional semiconductors and monolayers of transition metal dichalcogenides. 
The observed strong polaron-enhanced nonlinearity exists for both tetragonal and orthorombic phases of MAPbI$_3$, and remains stable at elevated temperatures. \newline

\end{abstract}

\begin{document}

\flushbottom
\maketitle

\thispagestyle{empty}

\section*{Introduction}

Excitons are solid state analogs of a hydrogen atom, appearing due to the Coulomb attraction between an electron in a conduction band and a hole in a valence band. In direct bandgap semiconductors they can be created optically by resonant absorption of photons. If the energy of exciton-photon interaction exceeds all characteristic broadenings in a system, the regime of the strong light-matter coupling is achieved. It is characterized by the emergence of hybrid half-light half-matter elementary excitations, known as exciton-polaritons. To drive a system into this regime, one needs to reach an efficient confinement of electromagnetic field, which can be achieved in semiconductor microcavities \cite{weisbuch1992observation}, optical waveguides \cite{Ciers2017}, photonic bound states in continuum \cite{kravtsov2020nonlinear} or leaky modes of photonic crystal slabs \cite{Gogna2019}. 
Exciton-polaritons demonstrate a set of remarkable properties, which makes them ideal candidates for both fundamental study of a variety of quantum collective phenomena~\cite{Ciuti2013} and modern optoelectronic applications~\cite{Liew2011}. In particular, from their photonic component they inherit extremely small effective mass and long decoherence time, while the presence of the excitonic component allows for the efficient polariton-polariton interactions, leading to the onset of the robust nonlinear optical response. The latter is experimentally revealed as a blueshift of a polariton line with the increase of a pumping power. 

From technological perspective, it is clearly highly desirable to have a material platform, which combines the thermal stability of excitons and polaritons with strong degree of their optical nonlinearity. For more than a decade, the study of the nonlinear excitonic response was focused on conventional semiconductor platforms including both narrow band gap (e.g. CdTe~\cite{kasprzak2006bose} and GaAs~\cite{bajoni2008polariton,gao2012polariton,nguyen2013realization,sturm2014all}) and wide band gap materials (e.g. GaN\cite{semond2005strong,marsault2015realization,liu2015strong} and ZnO~\cite{van2006exciton,li2013excitonic}), where excitons are well described 
by the hydrogen model. 

Essentially, while the blueshift per single polariton tells us about how well the nonlinear polariton systems performs at low excitation density, the maximum value of blueshift enabled by a polariton system characterises its nonlinear optical response at high  density. For wide bandgap materials excitons have much smaller effective size $a_B$ (exciton Bohr radius), which decreases the exciton-exciton interaction constant $V_{XX}$ (often referred as $g$, and defined mostly by the processes of electron and hole exchange \cite{ciuti1998role}), and thus the blueshift per polariton is reduced also. In the same time reduction of the Bohr radius substantially increases the Mott transition density. This, in general, allows to reach greater values of the maximal possible blueshift of an excitonic line.

The blueshift value can be evaluated as 

\begin{equation}
    \label{eq:EXtot}
    \Delta E_{\rm max} \approx V_{XX} n_{\rm max}/2 - |V_{XX2}| n_{\rm max}^2.    
\end{equation}
Here $n_{\rm max}$ is the inflection point of parabolic spectrum, defining the applicability range of Eq.~\eqref{eq:EXtot}, which is tentatively below the Mott transition density. 
The parameter $V_{XX2}$ corresponds to higher order correlations, which have an opposite sign and strongly suppress the blueshift at large densities \cite{emmanuele2020highly}.

This tendency is illustrated in Fig.~1a, where  linear scaling of $\Delta E_{max}$ with the exciton binding energy $E_B$ can be clearly seen for both the cases of bulk excitons (green line) and 2D quantum well excitons (orange line). 
It should be noted, that experimentally reported blueshifts are essentially
smaller, than the values given by this simple estimate. 
This is connected with 
practical problems of approaching the Mott transition limit without substantial heating of the sample and other undesired side effects.
The observed values of polaritonic blueshift for various materials are summarized in Table 1. Up to our knowledge, the maximal values reached so far are about 13 meV, observed in the microcavities with active media consisting on WS$_2$ monolayers.

The further details of  Coulomb interaction-induced exciton blueshift in different systems are presented in Section 4 of the \blue{Supplementary Material}.

\begin{figure}[t!]
\centering
\center{\includegraphics[width=0.9\linewidth]{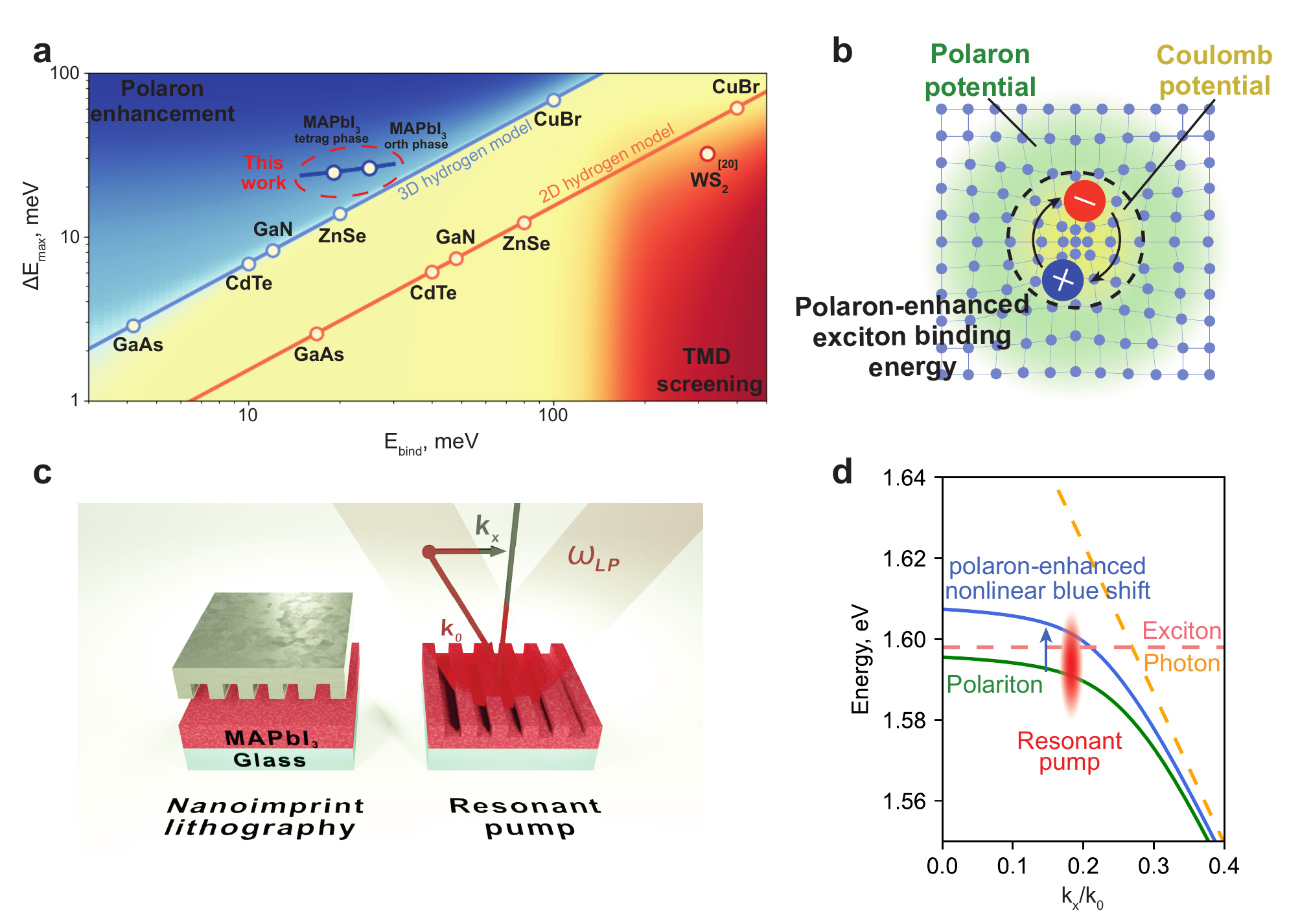}}
\caption{(a) The estimated values of maximal blueshift induced by exciton-exciton interaction in the quadratic approximation, $\Delta E_{\rm max}=V_{XX} n_{\rm max}/2 - |V_{XX2}| n_{\rm max}^2$, where $n_{\rm max} = 0.05/a_B^d$, $d=2,3$ is the density corresponding to maximal blueshift
for the 2D and 3D materials, respectively. 
The values corresponding to the hydrogen model, applicable to vast majority of the conventional semiconductors (GaAs, CdTe, GaN, ZnSe, CuBr are taken as examples) lie at straight blue (3D) and orange (2D) lines. The points below these line correspond to the materials, which are suboptimal for optical nonlinearity (e.g. TMD monolayers, red circle), the points above these line, such as MAPbI$_3$ perovskite considered in the present work (dark blue circles), to optimal materials.
The values of $n_{\rm max}$ for the exciton-polarons in perovskites and excitons in TMD monolayers are distinct from that of 3D and 2D hydrogen models, and depend on particular material parameters.   
(b) The sketch of a polaron-mediated exciton state, visualizing the polaron enhancement of exciton binding energy. 
The polaronic renormalization of Coulomb interaction breaks the Rydberg type scaling between exciton binding energy and Bohr radius, resulting in modification  of exciton optical nonlinearity.
(c) The sketch of the experimental geometry used in this work. Photonic crystal slab is fabricated by nanoimprint lithography method. The corresponding leaky photonic mode couples with excitonic transition in bulk MAPbI$_3$ perovskite giving rise to polariton modes, which are analyzed in nonlinear reflectance measurements performed under resonant (in angle and frequency) laser pump. The incident wavevector and its in-plane component are denoted by $k_0$ and $k_x$, respectively.   
(d) Calculated lower polariton dispersion (green line) resulting from the strong coupling between the exciton (light-red dashed line) and the leaky photonic crystal modes (yellow dashed line). 
The increase of the resonant pump leads to a blueshift of the lower polariton mode (solid blue line) caused by many-body renormalization of spectrum due to the exciton-exciton interactions and quenching of the Rabi splitting (see main text for corresponding discussion).
}
\label{fig:concept}
\end{figure}

The deviation from the discussed tendency is possible, if electron-electron interaction differs substantially from the Coulomb law. 
One of the notorious examples of such materials are monolayer transition metal dichalcogenides (TMD), where the dimensional reduction makes an exciton state different from that described by the hydrogen model\cite{chernikov2014exciton}. This results, among the rest, in the strong deviation (Fig.~\ref{fig:concept}a) of its nonlinear behavior from the general trend~\cite{shahnazaryan2017exciton,kravtsov2020nonlinear,stepanov2021exciton}. Note, however, that from the point of view of excitonic nonlinearity the case of TMD monolayers is sub-optimal, as corresponding points lie below the line describing the case of the quantum wells of conventional semiconductors.
Moreover, it is still challenging to find a way for boosting excitonic nonlinearity for 3D materials. 

In turn, it is well known that hybrid halide perovskites posses strong electron-phonon interaction, which is defined by high softness of their crystalline lattice~\cite{wright2016electron}, see Fig.~\ref{fig:concept}b.
Moreover, excitons in these materials are characterized by relatively high exciton binding energy and oscillator strength\cite{su2021perovskite}, as well as high defect tolerance\cite{huang2017lead}, which make perovskites highly prospective for studying room-temperature exciton-polariton dynamics and even Bose-Einstein condensation~\cite{su2018room,su2020observation,feng2021all,su2017room,fieramosca2019two,bouteyre2019room}. The analysis of the corresponding nonlinearities thus seems to be an important task.

Here we present a clear experimental evidence, supported by theoretical modelling, that exciton-phonon coupling leading to the formation of exciton-polarons in hybrid halide perovskites substantially modifies exciton-exciton interaction and allows to dramatically increase nonlinear optical response, which is characterized by record high value of the polariton blueshift up to 19.7 meV,   remaining robust at elevated temperatures (170 K).

\begin{table}[]
\centering
\begin{tabular}{|l|l|c|c|c|}
\hline
Material Group                                          & Material               & Reported blueshift [meV] & \multicolumn{1}{l|}{ Temperature [K]} & \multicolumn{1}{l|}{Reference}   \\ \hline
\multicolumn{1}{|c|}{\multirow{6}{*}{Semiconductor QW}} & \multirow{3}{*}{GaAs}& 0.6 & 8 & \citeonline{brichkin2011effect}
\\ \cline{3-5} 
\multicolumn{1}{|c|}{}                                  & & 0.8 & <10 & \citeonline{mukherjee2019observation}
\\ \cline{3-5} 
\multicolumn{1}{|c|}{}                                  & & 2.2 & <10 & \citeonline{estrecho2019direct}
\\ \cline{2-5} 
\multicolumn{1}{|c|}{}                                  & CdTe & 1.8 & 19 & \citeonline{kasprzak2006bose}
\\ \cline{2-5} 
\multicolumn{1}{|c|}{}                                  & ZnO & 6.0 & 300 & \citeonline{li2013fabrication}
\\ \cline{2-5} 
\multicolumn{1}{|c|}{}                                  & GaN & 7.5 & 300 & \citeonline{christopoulos2007room}
\\ \hline
\multirow{4}{*}{TMD}                                    & \multirow{2}{*}{MoSe$_2$} & 3.0 & 127 & \citeonline{stepanov2021exciton}                              \\ \cline{3-5} 
                                                        & & 5.0 & 7 & \citeonline{kravtsov2020nonlinear}
                                                        \\ \cline{2-5} 
                                                        & \multirow{2}{*}{WS$_2$} & 1.0 & 300 & \citeonline{zhao2021ultralow}
                                                        \\ \cline{3-5} 
                                                        & & 13.0 & 300 & \citeonline{barachati2018interacting}
                                                        \\ \hline
\multirow{3}{*}{Polymers}                                & BODIPY-G1 & 6.0 & 300 & \citeonline{yagafarov2020mechanisms} 
\\ \cline{2-5} 
                                                        & MeLPPP & 10.5 & 300 & \citeonline{zasedatelev2019room}
                                                        \\ \cline{2-5} 
                                                        & mCherry & 12.1 & 300 & \citeonline{betzold2019coherence}
                                                        \\ \hline
\multirow{4}{*}{Perovskites}                             & (PEA)$_2$PbI$_4$ & 8.5 & 300 & \citeonline{fieramosca2019two}
\\ \cline{2-5} 
                                                        & CsPbBr$_3$ & 9.5 & 300 & \citeonline{su2018room}
                                                        \\ \cline{2-5} 
                                                        & CsPbCl$_3$ & 9.5 & 300 & \citeonline{su2017room}
                                                        \\ \cline{2-5} 
                                                        &  \bf{MAPbI$_3$}  & {\bf 13.0} & 170 &  \bf{This work}  
                                                        \\ \cline{2-5} 
                                                        & \bf{MAPbI$_3$} & {\bf 19.7} & 6 & \bf{This work} 
                                                        \\ \hline
\end{tabular}
    \caption{The values of experimentally observed  blueshifts of exciton-polaritons in various materials, including conventional semiconductors (GaAs, CdTe, GaN, ZnO), TMD monolayers (MoSe$_2$, WS$_2$), organic materials (BODIPY-G1, MeLPPP, mCherry protein)  and hybrid perovskites ((PEA)$_2$PbI$_4$, CsPbBr$_3$, CsPbCl$_3$, MAPbI$_3$). The previously reported maximal values of about 13 mev correspond to optical microcavities with active media consisting on WS$_2$ monolayers. In this work we report the record high value of 19.7 meV for  MAPbI$_3$ perovskites.}
\end{table}
 

\section*{Results}
\subsection*{Fabrication of MAPbI$_3$ photonic crystal slab}

The big technological advantage of the use of halide perovskites for photonic applications~\cite{sutherland2016perovskite} is the variety of low-cost wet chemistry synthesis protocols to get high-quality thin films~\cite{dunlap2018synthetic}, for which additional nanostructuring can be routinely carried out by such well-developed and efficient methods as nanoimprint lithography~\cite{makarov2017multifold}. In particular, in order to achieve the strong exciton-photon coupling regime, instead of using vertical Bragg cavities to confine photons, one can realize a 1D photonic crystal slab by directly imprinting CH$_3$NH$_3$PbI$_3$ (MAPbI$_3$) film. Since perovskites have relatively high refractive index contrast with both air and glass substrate ($n_{MAPbI3} \approx 2.2 - 2.5$, $n_{glass} \approx 1.5$), the well-localized leaky modes supported by the grating are characterized by local field enhancement enabling their strong coupling with excitons possessing high optical oscillator strength. 

In the current work, thin and smooth films of MAPbI$_3$ were fabricated by a spin-coating method in a nitrogen dry box~\cite{makarov2017multifold}. The obtained films were patterned by means of nanoimprint lithography using periodic mold with a rectangular profile, as it is schematically shown in Fig.~\ref{fig:concept}c. The resulting MAPbI$_3$ photonic crystal slab is characterized by a thickness of 150~nm, period of $d_{period} = 750$~nm, groves depth of 70~nm and lateral fill factor $f = d_{ridge}/d_{period} = 0.33$ \blue{(See Section 1 of Supplementary Material)}. 

\subsection*{Strong light-matter coupling regime and exciton-polariton nonlinearity}
\begin{figure}[t!]
\centering
\center{\includegraphics[width=0.7\linewidth]{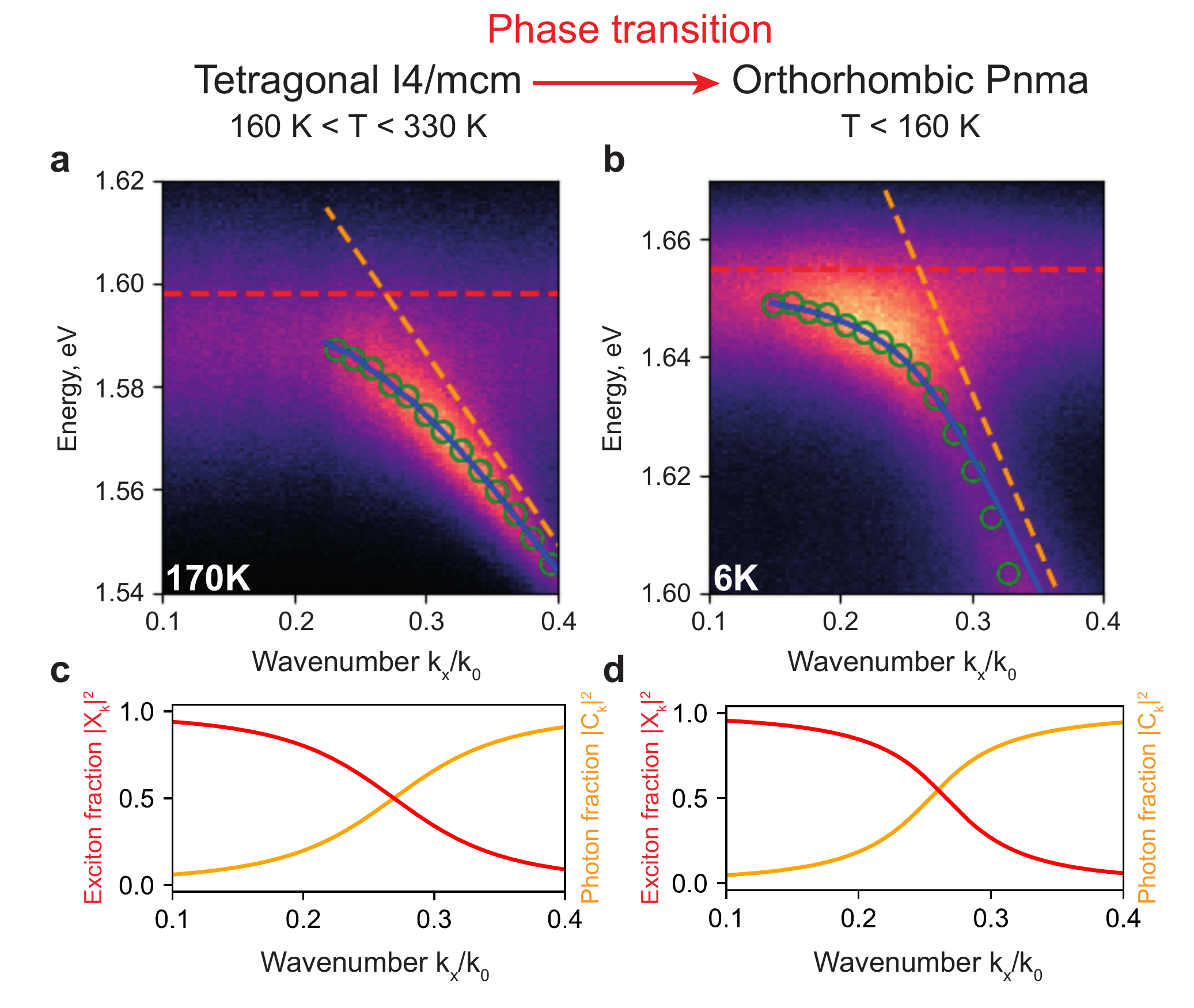}}
\caption{(a,b) Angle-resolved photoluminescence spectra of MAPbI$_3$ structured sample at 170K and 6K. Dashed yellow lines correspond to uncoupled photon cavity mode dispersions. Dashed red lines correspond to uncoupled exciton levels. Green empty dots denote extracted polariton dispersions. Blue solid lines correspond to the polariton modes fitted with the coupled oscillators model. (c,d) Hopfield coefficients $|X_k|^2$ for excitonic (red lines) and $|C_k|^2$ for photonic (yellow lines) fractions in polaritons calculated from coupled oscillator model.}
\label{fig2}
\end{figure}

To confirm the formation of exciton-polaritons, we performed angle-resolved photoluminescence (PL) spectroscopy measurements \blue{(see "Methods")} of the sample at T=170~K and T=6~K ( Fig.~\ref{fig2}a,b), i.e. above and below the temperature of MAPbI$_3$ phase transition ($\approx 160 K$) between tetragonal and orthorombic phases, respectively\cite{menendez2015nonhydrogenic, lee2021domain}. In both high-temperature tetragonal I4/mcm and low-temperature orthorombic Pnma phases, MAPbI$_3$ exhibits robust excitonic response, which manifests itself in the PL spectra as a broadband angle-independent peak at the energies 1598~meV and 1655~meV, respectively. As well, in both phases, the TE-polarized photonic mode exhibits clear anti-crossing with the corresponding exciton resonances. Fitting the dispersions of lower polariton branches (LPB) using coupled oscillator model~\cite{hopfield1958theory} (Fig.~\ref{fig2}a,b) allowed us to estimate the Rabi splitting $\Omega_{R} = 35.2$~meV and coupling strength $\Omega_0 = 35.8$~meV for the high-temperature tetragonal phase and $\Omega_{R} = 35.7$~meV and \ $\Omega_0 = 39.2$~meV for the low-temperature orthorombic phase (\blue{see "Methods" and Section 2 of Supplementary Material for the details}). The extracted Hopfield coefficients $|X_k|^2$ and $|C_k|^2$ representing, respectively, the angular dependence of excitonic and photonic fractions in the polaritons are shown in Fig.~\ref{fig2}c,d.  

It is well known that linear excitonic response in organic-inorganic lead halide perovskites is strongly affected by the polaron effects~\cite{soufiani2015polaronic, baranowski2020excitons,buizza2021polarons}. It thus becomes interesting to investigate how the polaron effects will modify polariton nonlinearities. To do so, we measure the blueshifts of polariton branches in both crystal phases under resonant femtosecond-pulsed excitation depending on pump fluence and excitonic fraction in the polariton mode, defined by the angle of incidence. 
In the experiment, the laser pump central frequencies and angles of incidence are chosen to resonantly probe each polariton branch with different excitonic fractions, as it is shown in Fig.~\ref{fig3}a,b. In each measurement, the spectral position of the polariton mode is extracted using Fano line shape fitting of the reflectance spectrum obtained within the spectrum of the pulse \blue{(see "Methods")}. 
Polariton mode spectra under resonance pump with fitted Fano line shape at the lowest detuning for 170K and 6K are shown in Fig.~\ref{fig3}c,d.
The resulting fluence-dependent polariton mode blueshifts obtained 
at T=170~K and T=6~K for various exciton fractions $|X_k|^2$ are shown in Figs.~\ref{fig3}e,f. The most pronounced blueshifts reaching to $\Delta E$=13~meV for T=170~K and $\Delta E$=19.7~meV for T=6~K are expectedly observed at smaller angles of incidence, corresponding to larger exciton fractions $|X_k|^2$. In both cases, the fluence dependence of the blueshifts is sub-linear, clearly demonstrating the effects of the saturation similar to those reported for TMD-based samples \cite{emmanuele2020highly}. 

The careful analysis of the linewidth of the polariton branch at  T=6~K reveals its broadening at $k_{||}/k_0 \approx 0.31$ \blue{(see Fig.~S5 in  Supplementary Material)}. We attribute this to the exciton resonance corresponding to the tetragonal phase, which coexists with the orthorhombic Pnma phase at T=6~K in polycrystalline thin film and is weakly coupled to the photons. It gives rise to resonant absorption which leads to the reduction of the pump efficiency and, thus, the magnitude of polariton blueshifts observed for $|X_k|^2$ = 0.158.

\begin{figure}[t!]
\centering
\center{\includegraphics[width=0.8\linewidth]{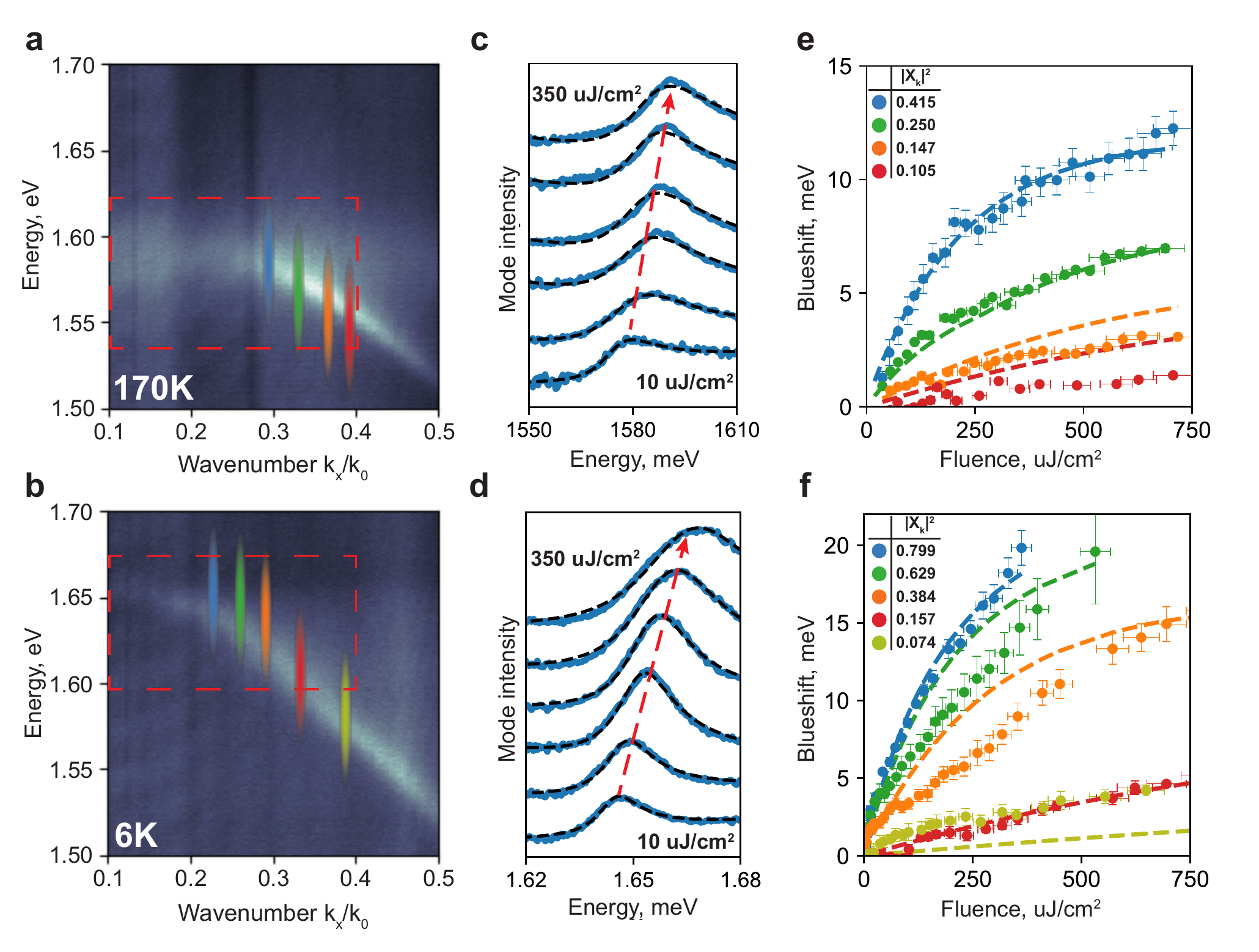}}
\caption{(a,b) Angle-resolved reflectance spectra of MAPbI$_3$ sample at 170K and 6K. Colored ellipses visualize angles and energies of resonant laser pump in the nonlinear reflectance measurements. Red dashed squares visualize the respective areas shown in Figs~\ref{fig2} (a,b). (c,d) Evolution of the reflectance spectra, shown as solid blue lines, with the increase of incident pump fluence under resonant femtosecond excitation of LPB, corresponding to the blue ellipses in (a,b) for 170K and 6K. Black dashed lines show Fano fits of the measured spectra.
(e,f) Extracted LPB spectral blueshifts under femtosecond laser excitation as a function of incident pump fluence for 170K and 6K. Corresponding theoretical calculations are shown with dashed lines. Inset tables show coefficients $|X_k|^2$ for excitonic fractions in polaritons. Dot colors correspond to ellipses colors at (a,b). Horizontal error bars represent the RMS deviation of the pump laser fluence. Vertical error bars correspond to the standard deviation error obtained from Fano lineshape fitting.}
\label{fig3}
\end{figure}
%


\section*{Discussion}

In order to unveil the physical origin of observed fluence-dependent blueshift of polariton spectrum, we develop a microscopic model of exciton-polariton response in the considered structure.
Our model includes the quantum mechanical description of exciton-polaron state in MAPbI$_3$ perovskite; the calculation of excitonic nonlinearity rates; the quantitative description of polariton gas temporal dynamics; the many-body renormalization of polariton resonance energy. 
As shown in Fig.~\ref{fig3}, the calculated spectrum demonstrates an excellent agreement with experimental data. We attribute the minor deviation observed for low excitonic fractions to the uncertainty in the estimation of the Hopfield coefficients, see Section 2 of \blue{Supplementary material}.

The excitonic properties of the considered material are strongly mediated by polaron effects, associated with coupling to longitudinal phonon (LO) mode \cite{baranowski2020excitons}.
Indeed, the  static dielectric permittivity of MAPbI$_3$ is $\epsilon_s = 25$ \cite{gelmetti2017selective}. In the 3D hydrogen-like model this corresponds to the exciton binding energy of about 3 meV, which makes excitons even less stable than in GaAs.
In direct contradiction to this, we experimentally found the resonance energies of the exciton transition as $E_X = 1598$ meV in the tetragonal phase, and $E_X = 1655$ meV at orthorhombic phase.
This is in perfect agreement with previous measurements \cite{soufiani2015polaronic}, where the corresponding exciton binding energies are reported as $E_b = 19$~meV, 25~meV, respectively. 
Fixing the high frequency dielectric permittivity $\varepsilon_\infty=5$, and altering the phonon energy, we reproduce these values within the  Pollmann-Büttner model for exciton-polarons \cite{pollmann1977effective, menendez2015nonhydrogenic}.
The resulting Bohr radius is of order of $a_B \sim 2.5$ nm. 

The nonlinear optical response in the regime of strong light matter coupling is governed by two effects. The first one is repulsive exciton-exciton exchange interaction, which shifts the position of the excitonic mode with increase of the pump. The reduction of the Bohr radius due to the polaronic effects discussed above leads to the decrease of overlap of excitonic wavefunctons and thus to the reduction of exciton-exciton Coulomb interaction calculated within the Born approximation \cite{ciuti1998role}. 
On the other hand, the small radius of the excitons boosts the density of Mott transition by 1-2 orders of magnitude as compared with conventional semiconductor materials, such as GaAs.
In turn, this allows to reach the regime of elevated particle densities, where the strong interparticle correlations are well pronounced, and result in particular in potentially giant  blueshifts (see. Fig.~\ref{fig:concept}a). 
Thus, the polaron-induced gain in the particle density overcomes the polaron-induced reduction of the Kerr nonlinearity per particle. The second mechanism of polariton nonlinearity is associated with the composite quantum mechanical statistics of excitons, which leads to the saturation of the optical absorption and corresponding quench of the Rabi splitting.
This mechanism of optical nonlinearity  is inherent for exciton-polaritons and is distinct from nonlinear response emerging in the domain of other types of polaritons, such as the vibrational polaritons \cite{f2018theory,ribeiro2018polariton,ribeiro2021enhanced}.
We also note, that the near-resonant excitation predominantly leads to formation of exciton-polaritons, with negligible free carrier concentration.
Therefore, we neglect the other third order nonlinear effects, related to collective response of free carriers  \cite{kalanoor2016third,ferrando2018toward}.

Both mechanisms can be treated on the equal footing in the coupled oscillator model, which results in the following expression for the blueshift as a function of the wavevector-resolved polariton density $n_{Lk}$, which can be estimated with use of the input-output formalism:

\begin{align}
    \label{eq:Epoltot}
    \Delta E_{LP}(k, n_{Lk}) \approx U (k)  n_{Lk} -  U_2 (k)  n_{Lk}^2 +\hat{O} (n_{Lk}^3) .
\end{align}
Here the expansion coefficients $U(k)$, $U_2(k)$ account for both Coulomb correlations and phase space filling effects and are calculated within the coboson formalism \cite{combescot2008many,emmanuele2020highly}. Their expressions are given in the \blue{"Methods"} section. We stress that the nonlinearity associated with the phase space filling scales as $a_B^{3}$ and in certain conditions is comparable with the Coulomb contribution (see Section 4 of \blue{Supplementary Material}). 
Notably, the quadratic in density term has an opposite sign as compared to the linear term, which leads to the saturation of the blueshift at elevated densities, as it is indeed in the experiment both for the tetragonal phase (Fig.~\ref{fig3}e) and the orthorombic phase (Fig.~\ref{fig3}f). 
In both cases, the reduction of blueshift with the growth of wavevector illustrated by Fig.\ref{fig3}a,b is caused by the corresponding reduction of the exciton fraction (see Fig.~\ref{fig2}c,d). 
For the orthorombic phase we report additional suppression of blueshift observed at intermediate values of wavevector (red and olive dots in Fig.~\ref{fig3}f). We attribute this effect to the residual fraction of the tetragonal phase, having an exciton with reduced oscillator strength lying about 50~meV lower in energy. 
The presence of this exciton plays a parasitic role, as it becomes unintentionally excited when pumping near its resonance. The further details of theoretical treatment are given in the \blue{"Methods"} section and \blue{Supplementary Material}.


\section*{Conclusion and outlook}

We have demonstrated  the formation of the robust nonlinear polariton response in the patterned bulk hybrid halide perovskite slabs. It has been shown that polaronic effects stemming from strong exciton-phonon interaction dramatically increase the stability of the excitons in perovskites, and enhance the corresponding nonlinear optical properties. The record high value of 19.7~meV for the polariton blueshift, which is about 50 percent higher than the values for other materials, has been reported. Experimental data are in good quantitative agreement with the results of the microscopic theoretical treatment. Our research opens the route to further exploration of the phonon-mediated polariton interactions for future energy efficient and thermally stable polaritonic devices.



\section*{Methods}

\subsection*{Sample fabrication}

\subsubsection*{Film synthesis}
The MAPbI$_3$ thin film was synthesised by solvent engineering method \cite{jeon2014solvent}. Solution of perovskite was prepared in the nitrogen dry box in the following way: 79.5~mg of methylammonium iodide (MAI) from GreatCell Solar and 230.5~mg of Lead(II) iodide (PbI$_2$) from TCI was dissolved in 1~ml DMF:DMSO solvent mixture in the ratio 9:1. Resulting 0.5M solution of MAPbI$_3$ was stirred for 1 day at 27 C$^{\circ}$. Film fabrication was performed in a nitrogen dry box by the spin-coating method. SiO$_2$ substrates (25$\times$25 mm) were washed with sonication in the deionized water, acetone and 2-propanol for 10~minutes consecutively, and afterwards cleaned in an oxygen plasma cleaner for 15~minutes. Perovskite solution (40~$\mu$l) was deposited on the substrate and spin-coated in one step at 4000~rpm for 40~sec. At 11 second 500~$\mu$L of toluene was dripped on the top of rotating substrate. Substrate with MAPbI$_3$ at intermediate phase was placed under vacuum for 3~minutes at room temperature to evaporate the residues of the solvents and toluene.
\subsubsection*{Nanoimprint lithography}
The resulting thin film was structured by the nanoimprint lithography method \cite{tiguntseva2019enhanced}. We used a DVD disk grating with the period 750~nm, 120~nm ridges height and the fill factor ($d_{ridge}/d_{period}$) 0.67 as a mold. The mold was cleaned in methanol and deionized water and then dried before the imprinting. The imprint process was carried out under 4.8~MPa for 10~minutes after the mold was removed. As the adhesion of SiO$_2$ substrates was quite large after the plasma cleaning, no antiadhesive layer was needed. Imprinted perovskite samples were annealed at 100 C$^{\circ}$ for 10~minutes. After the nanoimprint process was finished, the perovskite nanograting was formed with
negative replication of the used DVD disk mold.

\subsection*{Optical measurements}

The angle-resolved reflectance spectroscopy was performed using back-focal-plane imaging setup with a slit spectrometer coupled to a liquid-nitrogen-cooled imaging CCD camera (Princeton Instruments SP2500+PyLoN) and a halogen lamp employed for the white light illumination. Angle-resolved PL measurements were performed in the same setup with off-resonant excitation by monochromatic light from a femtosecond (fs) laser (Pharos, Light Conversion) coupled with a broad-bandwidth optical parametric amplifier (Orpheus-F, Light Conversion) and compressor at the wavelength of 680 nm, 40 fs pulse duration and 100 kHz repetition rate (see Supplementary information for details). For pump-dependent reflectivity measurements, the sample was excited by 40 fs pulses from the same wavelength-tuneable laser. The angle of incidence was controlled via focusing of the laser beam within the back focal plane of the objective lens. The sample was mounted in an ultra-low-vibration closed-cycle helium cryostat (Advanced Research Systems) and maintained at a controllable temperature in the range of 7-300K. The cryostat was mounted onto a precise XYZ stage for sample positioning. Spatial filtering in the detection channel was used to avoid parasitic signals originating from the reflections from the optical elements of the setup. More details on the experimental scheme are provided in \blue{ Supplementary Material Fig S4}.

\subsection*{Fitting of experimental data}

\subsubsection*{Fitting of polariton dispersions}
To extract the polariton dispersions at 170K and 6K from angle-resolved photoluminescence measurements, we first subtract the uncoupled exciton photoluminescence signal from the measured spectrum at each $k_x/k_0$  and fit the remaining data by Lorentz peak function. 
Combining peak positions for all $k_x/k_0$, we obtain the experimental polariton dispersion shown with empty green circles in Fig.~\ref{fig2}a,b.
The same method was used to fit the photon cavity dispersion at room temperature.
Though the upper polariton branch (UPB) cannot be observed due to the band absorption, the lower polariton branch can be fitted using coupled oscillator model \cite{hopfield1958theory}, with account for the spectral position and the linewidth of the uncoupled exciton: $\widetilde{E}_{X} = E_{X} - i\gamma_{X}$ and those parameters of the uncoupled cavity photon mode $\widetilde{E}_{C} (k) = E_{C} (k) - i\gamma_{C}$:
 
\begin{equation}
    E_{LP}^0 = \frac{\widetilde{E}_{X} + \widetilde{E}_{C}(k)}{2} - \frac{1}{2}\sqrt{(\widetilde{E}_{C}(k) - \widetilde{E}_{X})^2 + \Omega_0^2}, \label{CoupledOscillator}
\end{equation}
where $\Omega_0$ is the light-matter coupling strength. Resulting real part of $E_{LP}^0$ is shown in Fig.~\ref{fig2}a,b, imaginary part is shown in \blue{Supplementary Material Fig S5}. The Rabi 
splitting $\Omega_{R}$ corresponding to the spectral distance between LPB and UPB at the wavevector corresponding to the crossing of the uncoupled exciton and photon modes ($E_C = E_X$) 
can be calculated as

\begin{equation}
    \Omega_{R} =  \sqrt{\Omega_0^2 - (\gamma_{C} - \gamma_{X})^2 }.
\end{equation}
The temperature dependencies of the extracted values of $\Omega_{R}$, $\Omega_0$, $\gamma_{C}$ and  $\gamma_{X}$ are shown in Fig.~S12. It clearly reveals that for the studied system strong coupling criteria remain satisfied for the temperatures up to $230$~K.

\subsubsection*{Fitting of polariton blueshift under resonance pump}
To extract the nonlinear polariton blueshifts, we measured the angle-resolved reflectance spectra in TE and TM polarization under resonant femtosecond laser pump for several detunings (visualized as colored ellipses in Fig.~\ref{fig3}a,b). For each detuning, the reflectance spectrum in TM polarization was used as the reference, since no leaky polaritonic modes are supported in this polarization. By normalizing the TE-polarized spectrum by the TM-polarized reference for each fluence, we obtained the resonant Fano-line profiles corresponding to the lower polariton branches (see Fig.~\ref{fig3}c,d). The changes of central frequency as functions of incident fluence, describing polariton spectral blueshifts, are shown in Fig.~\ref{fig3}e,f. The linewidths of fitted polariton modes under resonance pump are provided in \blue{Supplementary Material Fig S6}.

\subsection*{Model} 

\subsubsection*{Exciton-polarons} 
Due to the polar nature of metal halide perovskites, their exciton response is governed by polaron effects \cite{baranowski2020excitons}.
Their presence primarily results in the longitudinal optical (LO) phonon-induced renormalization of the single-particle bandgap, the electron and hole effective masses, and the Coulomb interaction. 
The coupling between the phonons and the charged particles are characterized by the dimensionless Fröhlich coupling constants:

\begin{equation}
    \alpha_{e[h]} = \frac{e^2}{4\pi\varepsilon_0} \frac{1}{\hbar \varepsilon^*} \sqrt{\frac{m_{e[h]}}{2E_{LO}}} ,
\end{equation}
where $m_{e[h]}$ stands for bare electron [hole] mass, $E_{LO}$ is the energy of LO phonon,

\begin{equation}
    \frac{1}{\varepsilon^*} = \frac{1 }{\varepsilon_{\infty}}
    - \frac{1}{\varepsilon_s},
\end{equation}
with $\varepsilon_s$, $\varepsilon_\infty$ being the static and high frequency dielectric constants, respectively. 
We employ the Pollmann-Büttner model \cite{pollmann1977effective, menendez2015nonhydrogenic} for exciton-polarons, solving the Schrodinger equation

\begin{equation}
    \left[-\frac{\hbar^2 \nabla^2}{2\mu^*} - V(\vec{r}) \right] \psi(\vec{r}) = -E_b \psi(\vec{r}),
\end{equation}
where

\begin{equation}
    \label{eq:Vr}
    V(r) =  \frac{e^2}{4\pi \varepsilon_0 r} \left[ \frac{1}{\varepsilon_s } +\frac{1}{\varepsilon^* } 
    \left( \frac{m_h}{\Delta m} e^{-\frac{r}{l_h}} - \frac{m_e}{\Delta m} e^{-\frac{r}{l_e}}\right) \right]
\end{equation}
is the dressed Coulomb interaction, and $\mu^*=m_e^* m_h^*/ (m_e^*+m_h^*)$ is the exciton reduced mass. Here $e$ is the elementary charge, $\varepsilon_0$ is the vacuum permittivity,
$m_{e[h]}^* = m_{e[h]} \left(1+ \alpha_{e[h]}/6 \right)$, $\Delta m = m_h - m_e$, and $l_{e[h]} = \hbar /\sqrt{E_{LO} m_{e[h]} } $
are the polaron radii. 

For numerical calculations, we use the values presented in the review article \cite{baranowski2020excitons}. 
We set $\varepsilon_{\infty} = 5$,
$\varepsilon_{s} = 25$,
$m_e = 0.19 m_0$, $m_h = 0.22 m_0$,
where $m_0$ is the free electron mass. 
The value of LO phonon energy $E_{LO}^{T}=22$ meV is chosen to reproduce the previously reported \cite{soufiani2015polaronic} exciton binding energy $E_b^{T}\approx 19$ meV at $T=170$ K for tetragonal ($T$) phase. 
The resulting exciton wave function is of hydrogenic shape, and is well fitted with $\psi = \frac{1}{\sqrt{\pi a^3}} e^{-r/a_B}$, where $a_B = \langle \psi |r|\psi \rangle$ is the Bohr radius. 
The corresponding value of the Bohr radius is $a_B^{T}=2.36$ nm. 
For the low temperature region $T<10$ K in orthorombic ($O$) phase $E_b^{O} =25 $ meV is reported \cite{soufiani2015polaronic}, which corresponds to  $E_{LO}^{O}=16$ meV. 
The resulting Bohr radius is $a_B^{O}=2.14$ nm.
The dependence of excitonic properties on the energy of LO mode and the temperature is presented in \blue{Supplementary Material Fig S9}.

\subsubsection*{Nonlinearity coefficients}
The presence of the polaron effects makes
MAPbI$_3$ a unique platform for probing collective phenomena associated with \textit{bulk} exciton-polaritons. 
Indeed, the exciton binding energy of about 20-25 meV and the Bohr radius of $a_B \sim$ 2.5 nm  allows the operation at elevated temperatures and high particle densities. 
Supplemented by pronounced interparticle correlations, this allows to reach
large values of the blueshift of polariton energy.
The nonlinear spectrum of the lower polariton branch is described by Eq. \eqref{eq:Epoltot}, which is obtained as Taylor decomposition of the Eq.\eqref{CoupledOscillator} in polariton concentration $n_{Lk}$ up to the second order, thus accounting for the two-body and three body correlations.  Both Coulomb exchange interactions between excitons and reduction of the Rabi splitting due to phase space filling effects were accounted for within the so-called coboson diagrammatic technique, developed by M. Combescot with co-authors \cite{combescot2008many}. 
The resulting nonlinear coefficients read
(see Ref.~\cite{emmanuele2020highly} for the derivation):

\begin{align}
    U (k) \approx 
    \frac{V_{XX}}{2} |X_k|^4 
    + \frac{\Omega_0 }{2} s |X_k|^2 (X_k^* C_k + X_k C_k^*),
\end{align}
\begin{align}
    U_2 (k) \approx 
    |V_{XX2}| |X_k|^6 
    + \frac{\Omega_0 }{2} |s_2| |X_k|^4 (X_k^* C_k + X_k C_k^*),
\end{align}
We accounted for only the primary contribution of Coulomb interactions, corresponding to zero exchange momenta \cite{carusotto2013quantum}.
The explicit expressions for the calculation of Coulomb correlations $V_{XX}$, $V_{XX2}$ and the saturation rates $s$, $s_2$ are given in \blue{Supplementary Material}. 
We note that due to hydrogenic shape of exciton wave function the saturation factors are computed analytically and give $ s = 7\pi a_B^3 $, $s_2 = - 253 \pi^2 a_B^6 /4$. 
The calculated values of exciton-exciton Coulomb scatterings are $V_{XX}^{T} = 0.031$ $\mu$eV$\cdot \mu$m$^3$, $V_{XX2}^{T} = -2.5 \cdot 10^{-9}$ $\mu$eV$\cdot \mu$m$^6$.
At low temperature we get $V_{XX}^{O} = 0.029$ $\mu$eV$\cdot \mu$m$^3$, $V_{XX2}^{O} = -2 \cdot 10^{-9}$ $\mu$eV$\cdot \mu$m$^6$.

\subsubsection*{Dynamics}
We estimate the polariton density within the input-output formalism.
In the lowest order of the mean field approximation the polariton density is $n_{Lk} = |\bar{p}_k|^2$, where 
$\bar{p}_k = \langle \hat{p}_{Lk} \rangle$, and $\hat{p}_{Lk}$ denotes the polariton annihilation operator.
The dynamic equation for $\bar{p}_k$ reads:

\begin{align}
    \label{eq:pdyn}
    \hbar \dot{\bar{p}}_k = & - i E_{LP} (k) \bar{p}_k
    - \frac{ \gamma_{L}(k) + \gamma^\prime}{2} \bar{p}_k 
    -|X_k|^4 \gamma_2 |\bar{p}_k|^2 \bar{p}_k  
    -  2 i U(k)|\bar{p}_k|^2 \bar{p}_k    
    +  3 i U_2(k) |\bar{p}_k|^4 \bar{p}_k 
    + \beta_k \sqrt{\frac{\gamma_L (k)}{2} } a_0 e^{-t^2/(2\tau_p^2)} e^{-i E_Lt/\hbar}
\end{align}
Here $\gamma_L (k)$ [$\gamma^\prime$] are the radiative [non-radiative] decay rates of lower polariton branch, $\tau_p$ is the pulse duration. $a_0$ is the
square root of the number of photons passing through the structure per unit time per unit
area, which is related to the peak incident power density $a_0 = \sqrt{F/(\omega_{LP} \tau_p L_C )}$, with $F$ being the pump fluence, $L_C$ denoting the sample length.
The parameter $\gamma_2$ is the exciton-exciton annihilation rate. 

We numerically simulate the Eq.~\eqref{eq:pdyn} and assume that the optical response is collected at maximal density, $n_{Lk} = n_{Lk}(t)|_{\rm max}$.
The values of the parameters are given in the \blue{Supplementary Material}.
The resulting blueshifts of the polariton energy at $T=170$ K calculated via Eq.~\eqref{eq:Epoltot} are shown in Fig.~\ref{fig3} a, demonstrating a good agreement with experimental evidence.
The only fitting parameter in our model is the exciton-exciton annihilation rate, which is fixed at $\gamma_2^{T} = 3.75 V_{XX}^{T}$.

In order to explain the behavior of the polariton blueshifts at $T=6$ K, we assume the presence of rudimentary tetragonal phase, clearly visible in the luminescence spectrum (see the \blue{Supplementary Material}).
This results in unintentional pumping of the corresponding weakly coupled exciton mode at intermediate values of wavevector. We treat such mode as an exciton reservoir. For the sake of simplicity  we neglect the reservoir-induced nonlinear effects. 

The full treatment would contain numerous cross-nonlinear terms between the excitons from different phases, thus overburdening the model. 

Yet, we quantify the presence of the reservoir introducing the parameter $\beta_k = \gamma_L(k)/[\gamma_{X0}(k) + \gamma_L(k)]$, 
which defines the efficiency of the pumping of the polariton branch. The exciton linewidth $\gamma_{X0}(k)$ is extracted via decomposition of the pure polariton linewidth (see the \blue{Supplementary Material} for the details).
The resulting simulation of the blueshift is shown in Fig.~\ref{fig3} b, with $\gamma_2^{O} = 2 V_{XX}^{O}$ and demonstrates good agreement with experiment at low wavevectors, together with reduced blueshift at higher wavevectors.
We attribute the latter effect to the neglected contribution of the reservoir induced  nonlinearity.

We attribute the broadening of the linewidth with increasing pump fluence to  exciton-exciton annihilation.
In principle such inhomogeneous broadening should be treated in self-consistent manner \cite{ciuti1998role}.
However, one can approximate the growth of linewidth phenomenologically as \cite{gribakin2021exciton}:

\begin{equation}
   \Delta\gamma = \gamma_2 \frac{n_{Lk}}{\sqrt{1+n_{Lk}/n_1}}.
\end{equation}
where $n_1 = 0.5 \cdot 10^{19}$ cm$^{-3}$ is a variationally optimized parameter.
The corresponding experimental data and theoretical fits are shown in \blue{Supplementary Material}.

\section*{Acknowledgements}

The experimental part of this work was funded by Russian Science Foundation, grant \#21-12-00218.
The theoretical part of this work carried out by V.A.S., I.V.I, and I.A.S. was partially supported by SC RA project number SCS 20RF048. I.A.S. acknowledges as well the support of the Icelandic Research Fund (Rannis), project No. 163082-051.


\section*{Data availability}
The data that support the findings of this study are available from the authors upon reasonable request, see author contributions for specific data sets.


\section*{Competing interests}
The authors declare no competing interests.

\section*{Author contributions statement}
M.A.M. conducted the experiments. V.A.S. developed and implemented the model. M.A.M. and V.A.S. fitted the experimental data. F.A.B. provided experimental assistance. I.A.S. and I.V.I. supervised the theoretical work. S.V.M., D.N.K., and A.K.S. supervised the experimental work. A.K.S. managed the project. All authors extensively discussed the results and  participated in editing of the manuscript.

\section*{Table of Contents}

In this work, exciton-polaritons in photonic crystal slab based on halide perovskite are studied. It revealed the influence of the polaron effects on the exciton state and polariton interaction. As a result, it is experimentally observed a record-high value of the nonlinear optical response detected as a 19.7 meV blueshift of the polariton branch under femtosecond laser irradiation.

\begin{figure}[t!]
\centering
\center{\includegraphics[width=0.8\linewidth]{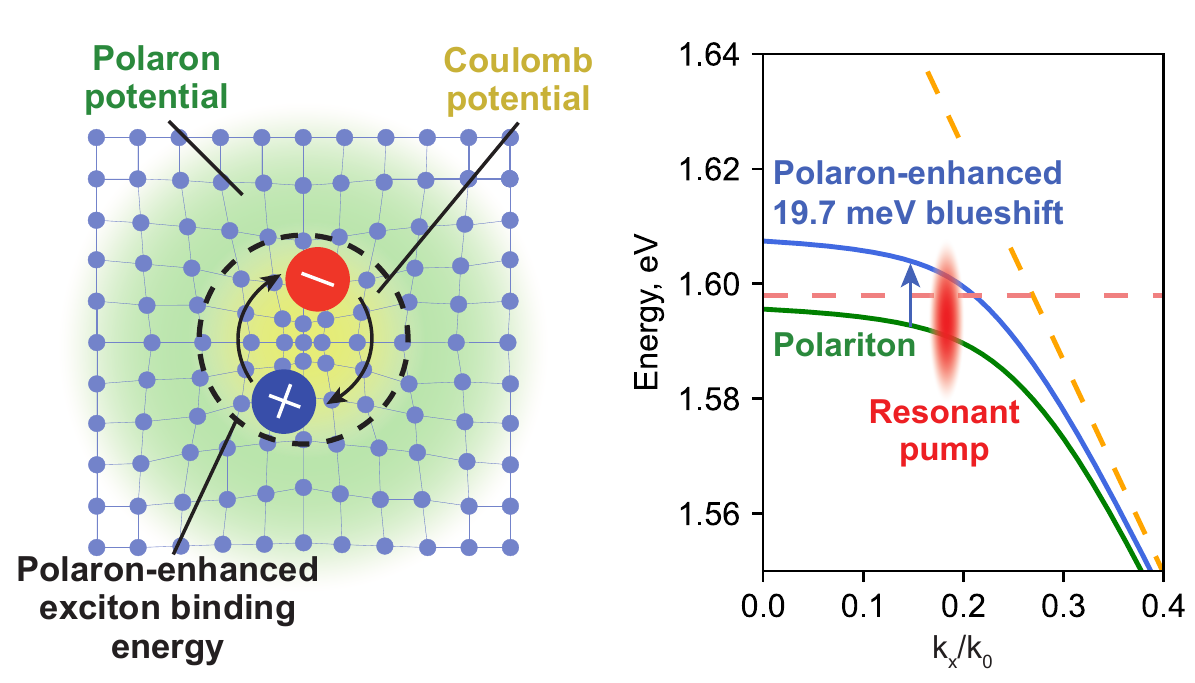}}
\caption{ToC figure}
\label{fig3}
\end{figure}

\clearpage

\setcounter{figure}{0}
\setcounter{equation}{0}

{ \huge \bf \noindent
Supplementary material: Polaron-enhanced polariton nonlinearity in lead halide perovskites
}

\vskip 15pt

{\large \bf \noindent 
Mikhail~A.~Masharin$^{1,\dagger}$,
Vanik~A.~Shahnazaryan$^{1,\dagger}$,
Fedor~A.~Benimetskiy$^{1}$,
D.~N.~Krizhanovskii$^{1,2}$,
Ivan~A.~Shelykh$^{1,3}$,
Ivan~V.~Iorsh,$^{1}$,
Sergey~V.~Makarov$^{1}$,
Anton~K.~Samusev$^{1,*}$
}

\vskip 15pt

\noindent $ ^1$ ITMO University, School of Physics and Engineering, St. Petersburg, 197101, Russia

\noindent $ ^2$ Department of Physics and Astronomy, University of Sheffield, Sheffield S3 7RH, United Kingdom

\noindent $ ^3$ Science Institute, University of Iceland, Dunhagi 3, IS-107, Reykjavik, Iceland

\noindent $ ^*$ Corresponding author: anton.samusev@gmail.com

\noindent $ ^\dagger$ These authors contributed equally: Mikhail A.~Masharin, Vanik A.~Shahnazaryan

\vskip25pt%
 
\section{Characterization of the sample: SEM, AFM and angle-resolved spectroscopy at room temperature}

Fig. \ref{fig:S1}a shows scanning electron microscopy (SEM) image of nanostructured MAPbI$_3$ film, which precisely reveals the period of the grating being equal to 750 nm. The topographic profile of the sample obtained using atomic force microscopy (AFM) measurements (Fig. \ref{fig:S1}b) gives the width (250 nm) and the height (85 nm) of the imprinted combs. The overall film thickness measured by AFM equals 150 nm. 

To initially characterize the optical properties if the imprinted MAPbI$_3$ sample, we have measured and simulated its angle-resolve reflectivity (Fig. \ref{fig:S1}c) at room temperature. The simulations were carried out using   Fourier Modal Method (FMM)\cite{li1997new} with account for the MAPbI$_3$ complex dielectric permittivity measured by means of ellipsometry.  The modelling results nicely agree with the experimental data and allow to extract the dispersion of the leaky photonic mode uncoupled from exciton resonance at room temperature. Note that the absorption at the frequencies above 1.63 eV  leads to the broadening and attenuation of the photon mode.

\section{Extraction of the uncoupled cavity photon mode dispersion}

\begin{figure}
\centering
\center{\includegraphics[width=0.4\linewidth]{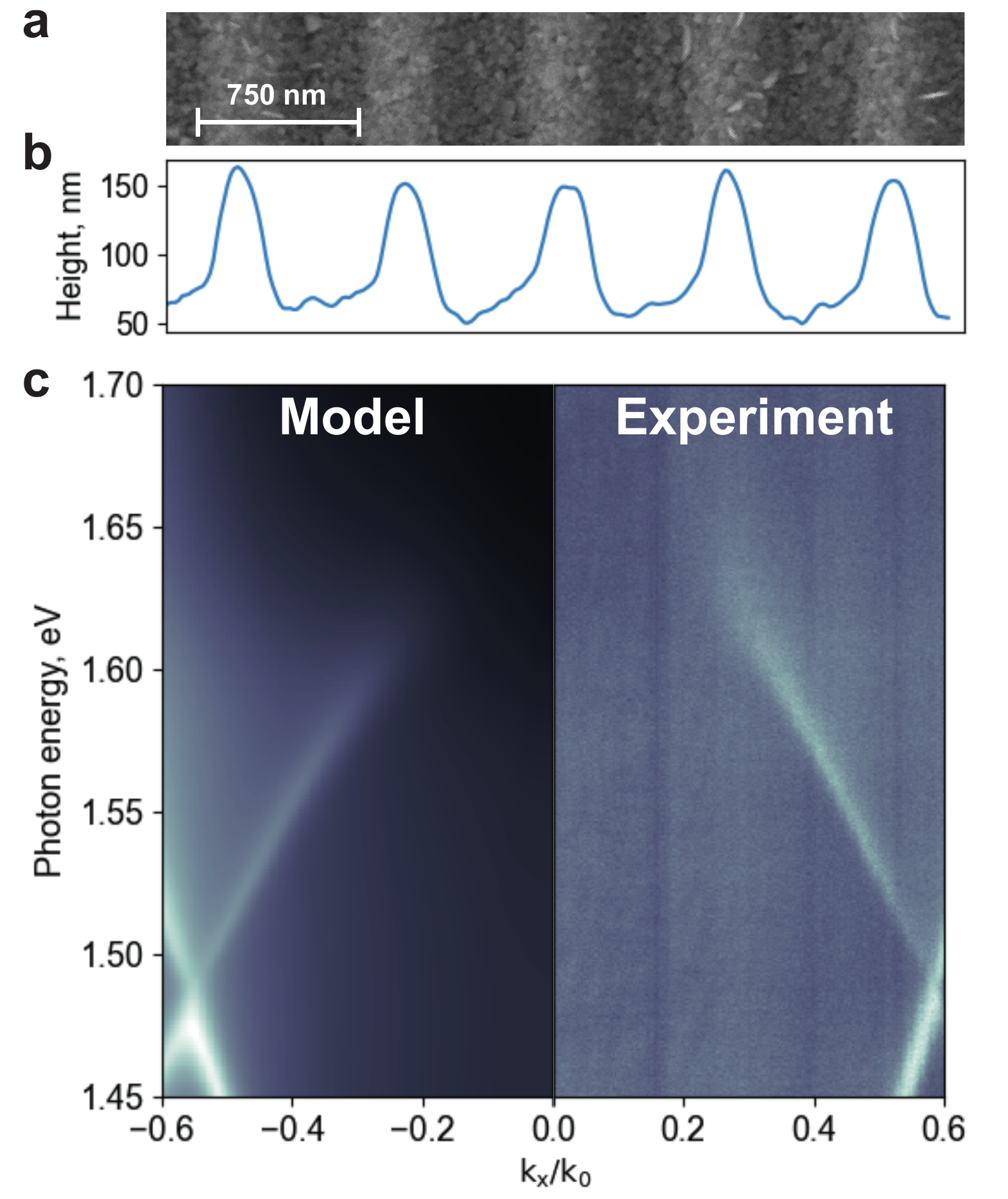}}
\caption{(a) SEM image of the structured MAPbI$_3$ film, scale bar is 750 nm. (b) AFM profile measured along the periodicity of the grating imprinted in the sample. (c) Simulated and experimentally measured angle-resolved reflectance spectra at room temperature}
\label{fig:S1}
\end{figure}

\begin{figure}
\centering
\center{\includegraphics[width=0.7\linewidth]{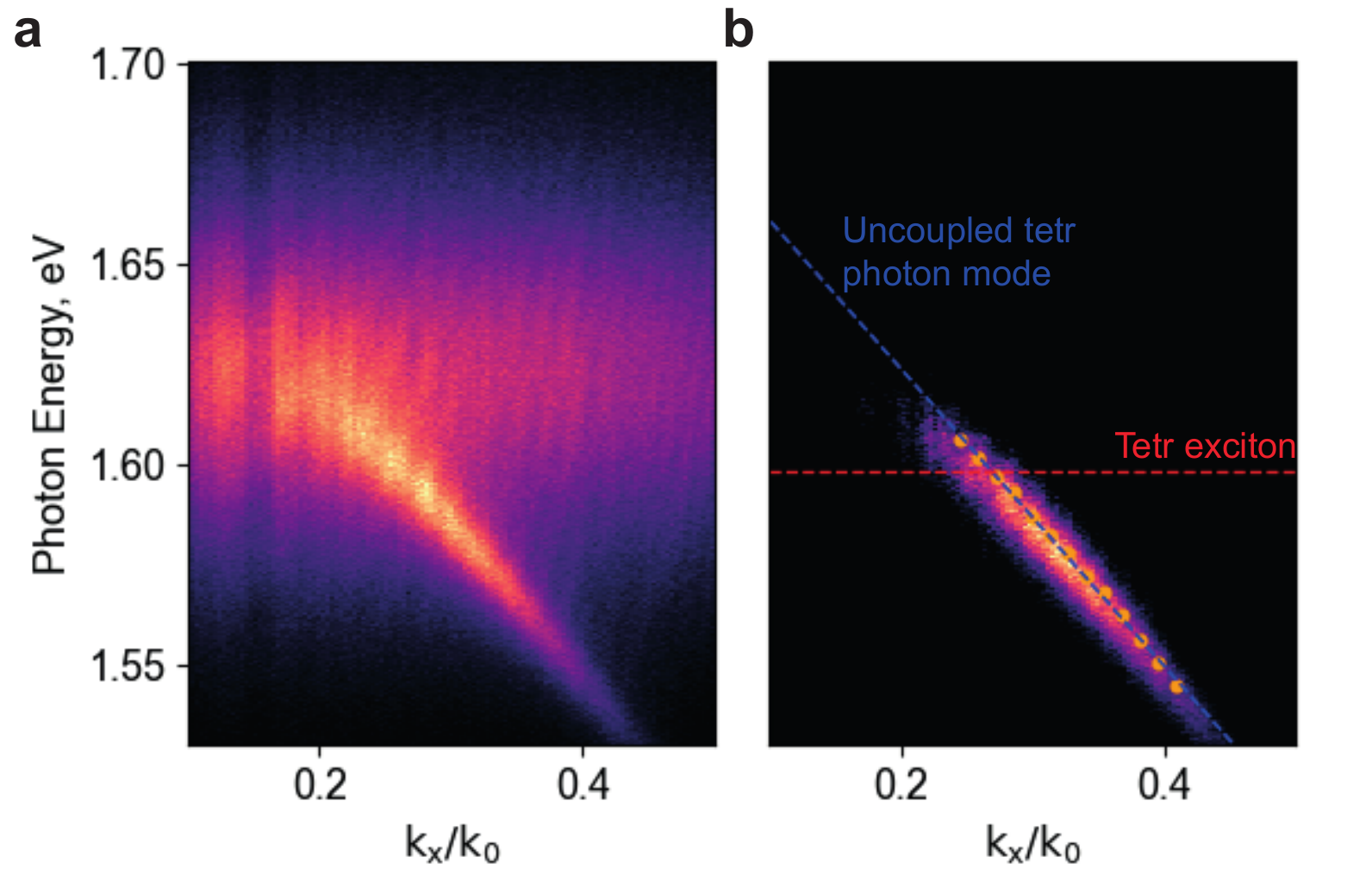}}
\caption{(a) Angle-resolved PL spectrum at room temperature and (b) extracted photon cavity mode from angle-resolved PL, obtained by subtraction of TM-polarized angle-resolved PL map from the TM-polarized one. Blue dashed line is the uncoupled photon cavity mode extrapolated from the in the spectral region far from exciton resonance for 170K crystalline phase. Red dashed line shows the energy of the exciton at 170K.}
\label{fig:S2}
\end{figure}

To estimate the strong exciton-photon coupling parameters (Rabi spitting, Hopfield coefficients etc.), along with the measured exciton-polariton dispersion, the uncoupled exciton energy and uncoupled photon cavity dispersion are needed. The exciton energy revealed from photoluminescence (PL) and reflectivity measurements is in excellent agreement with previously reported data\cite{soufiani2015polaronic1}. Meanwhile, the dispersion of the uncoupled photon cavity mode without exciton contribution is not obvious to reveal in the studied system. As shown in Fig \ref{fig:S1}c, we are unable to measure full photon cavity mode dispersion due to the absorption. Moreover, even at room temperature exciton contribution is noticeable, which can be seen from ellipsometry measurements \cite{jiang2016temperature} and becomes clear from our the FMM modelling and experiment (Fig \ref{fig:S1}c). Therefore, uncoupled photon cavity mode was linearly extrapolated from the measured angle-resolved reflectance spectra at large k$_x$, where exciton contribution is negligible. In order to get rid of the uncoupled exciton peak, we first subtracted the angle-resolved PL spectral map obtained in TM polarization from the one measured with TE-polarized analyzer. Further on, we have fitted the photon cavity peaks at each k$_x$ by Lorentz resonant function, shown in Fig \ref{fig:S2}. Since in the absence of exciton there is no disturbance of the photon cavity dispersion, we used linear fit to extend it up to the frequency of exciton transition at 170K. 

\begin{figure}
\centering
\center{\includegraphics[width=0.9\linewidth]{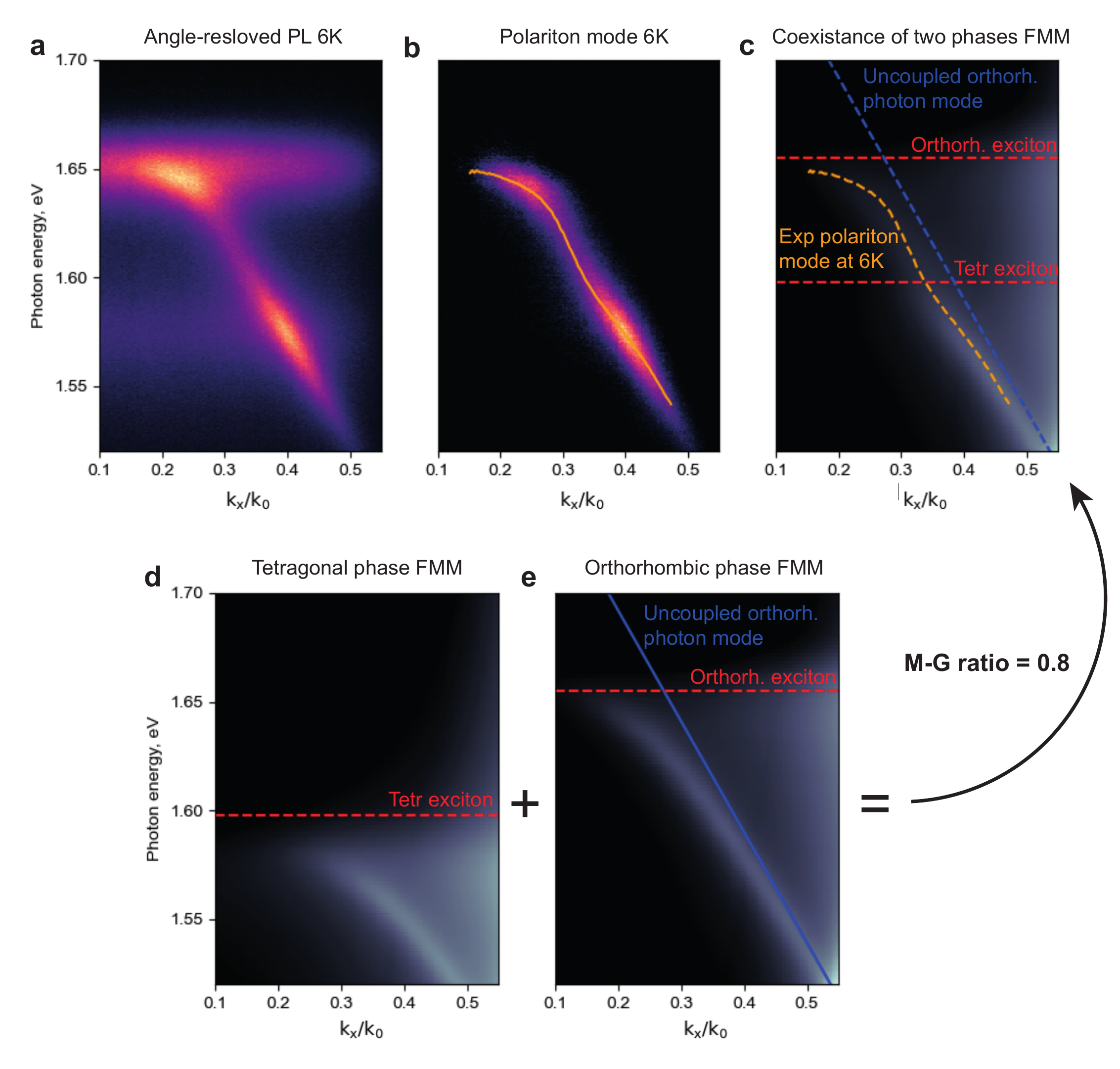}}
\caption{(a) Measured angle-resolved PL spectrum at 6K. (b) Extracted photon cavity mode at 6K obtained by subtraction of TM-polarized PL spectrum containing only the uncoupled exciton resonance. Yellow solid line shows extracted polariton mode. (c) Angle-resolved reflectance spectrum calculated using FMM for two coexisting crystalline phases, estimated using Maxwell-Garnet approximation with the ratio of 0.8. Red dashed lines corresponds to the levels of exction resonances at 170K and 6K. Blue line represents estimated uncoupled photon cavity mode for orthorhombic phase. Simulated angle-resolved reflectance spectra for (d) pure tetragonal and (e) pure orthorhombic phases.}
\label{fig:S3}
\end{figure}

Due to the phase transition and coexistence of two crystal phases in polycrystalline film at 6K mentioned in the main text, the uncoupled photon cavity mode of orthorhombic MAPbI$_3$ phase is affected by the exciton resonance of tetragonal one. As discussed in Ref.~\citeonline{lee2021domain1}, tetragonal and orthorhombic phases in polycrystalline MAPbI$_3$ film coexist as a core-shell-type perovskite grains at low temperatures. The core has the orthorhombic phase and the shell represents tetragonal phase. To obtain effective dielectric permittivity of this system we used Maxwell-Garnet (M-G) approximation\cite{markel2016introduction}. Accounting for the refractive index of the materials from two crystal phases measured previously\cite{jiang2016temperature}, we vary the ratio between two phases in effective refractive index, calculated with M-G approximation to match the simulated angle-resolved reflectance spectrum with the experimental one. Resulting effective refractive index corresponds to a ratio between two crystal phases of 0.8 (This corresponds to 80\% of orthorhomic phase and 20\% of tetragonal phase). The fitting results are shown in Fig \ref{fig:S3}. In order to estimate the uncoupled photon cavity dispersion at 6K, we simulate the angle angle-resolved reflectance spectrum for only the orthorhombic phase and linearly extrapolate the dispersion from the spectral region far from the exciton resonance to higher frequencies (Fig \ref{fig:S3} e). In further analysis of the polariton mode dispersion, this linearly fitted cavity photon mode was considered.

\begin{figure}
\centering
\center{\includegraphics[width=0.8\linewidth]{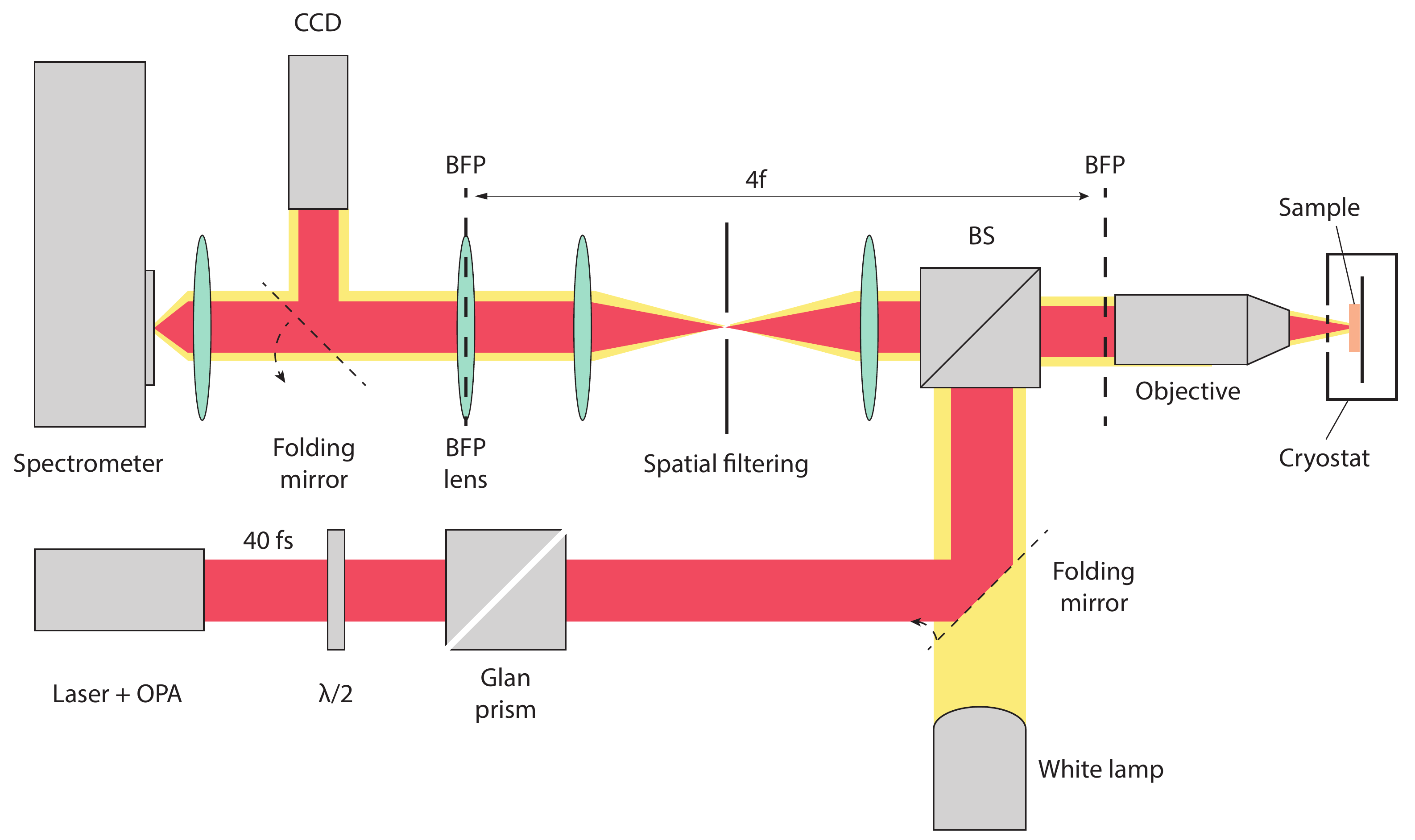}}
\caption{Scheme of experimental setup for angle-resolved reflectivity and photoluminescence measurements. White halogen lamp is used for the measurements of linear optical response, while femtosecond laser coupled with an optical parametric amplifier is used as a non-resonant pump in photoluminescence measurements and as a resonant pump in the nonlinear experiments. In order to obtain angle-resolved spectra, back focal plane (BFP) imaging was combined with the real-space filtering in a 4f configuration. The latter allows to eliminate unwanted background signal. Sample is placed in a closed-cycle He-free cryostat with micrometric positioning along 3 spatial axes. Spatially resonant laser excitation is achieved by focusing laser beam into the BFP of the objective lens granting 43~um laser spot at the sample surface. To vary the angle of incidence of the laser beam on the sample surface, two mirrors are used to realize parallel displacement of the laser beam and its focus within the BFP of the objective lens. The reference spectrum of the laser pulse is  in recorded in TM polarization, where any resonant features in the region of interest are absent. All signals was detected with an imaging Princeton Instruments spectrometer (f = 500 mm, 600 g/mm) and liquid nitrogen cooled PyLoN eXcelon CCD camera.}
\label{fig:S4}
\end{figure}

\begin{figure}
\centering
\center{\includegraphics[width=0.7\linewidth]{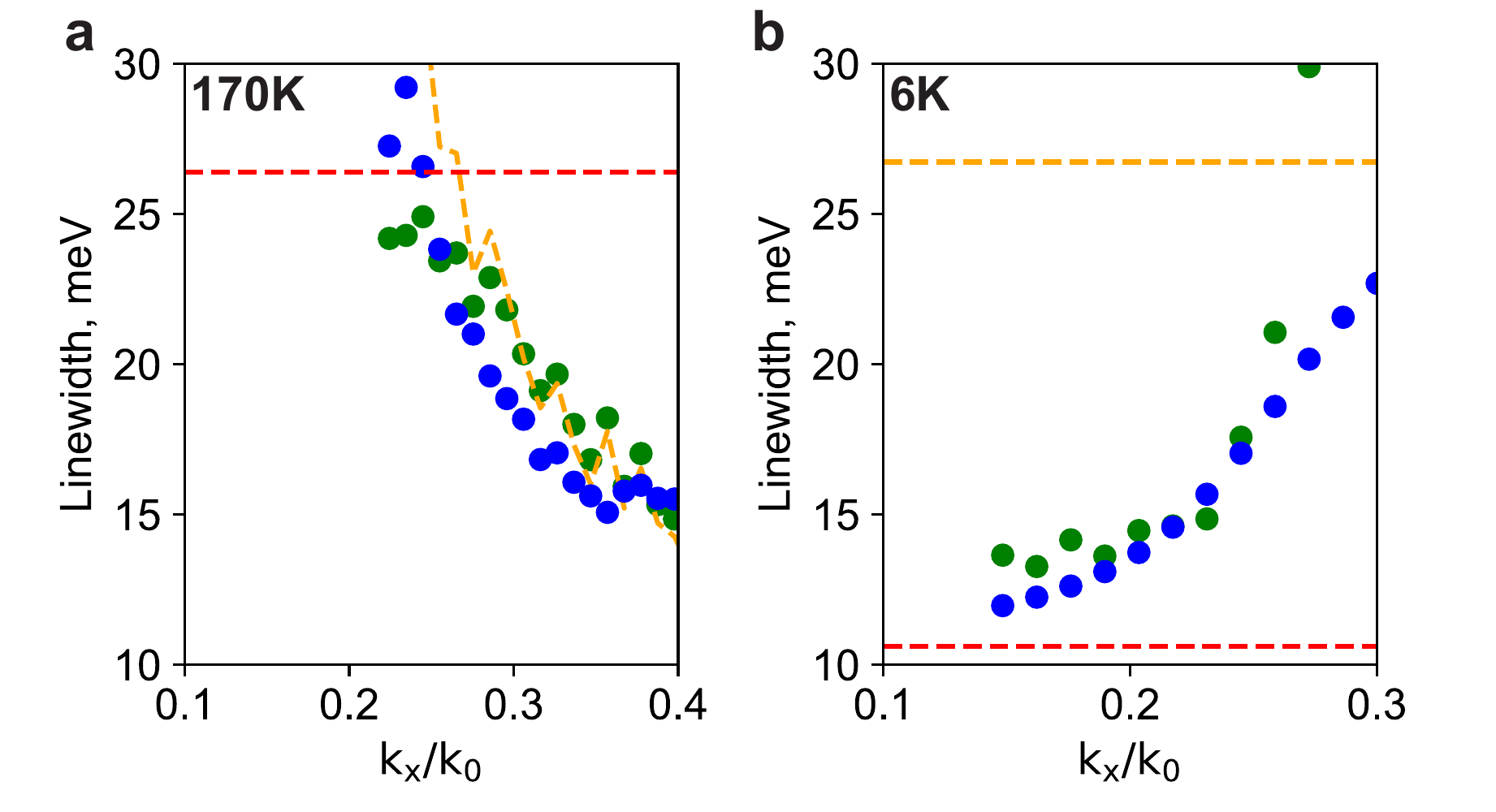}}
\caption{Linewidth as a function of in-plane wavevector of measured photon cavity mode dispersion (yellow dashed line), exciton resonance energy (red dashed line), polariton mode (green dots) and polariton mode fitted by two-coupled oscillator model (blue dots) at 170K (a) and at 6K (b).}
\label{fig:S Rabi}
\end{figure}

\begin{figure}
\centering
\center{\includegraphics[width=0.8\linewidth]{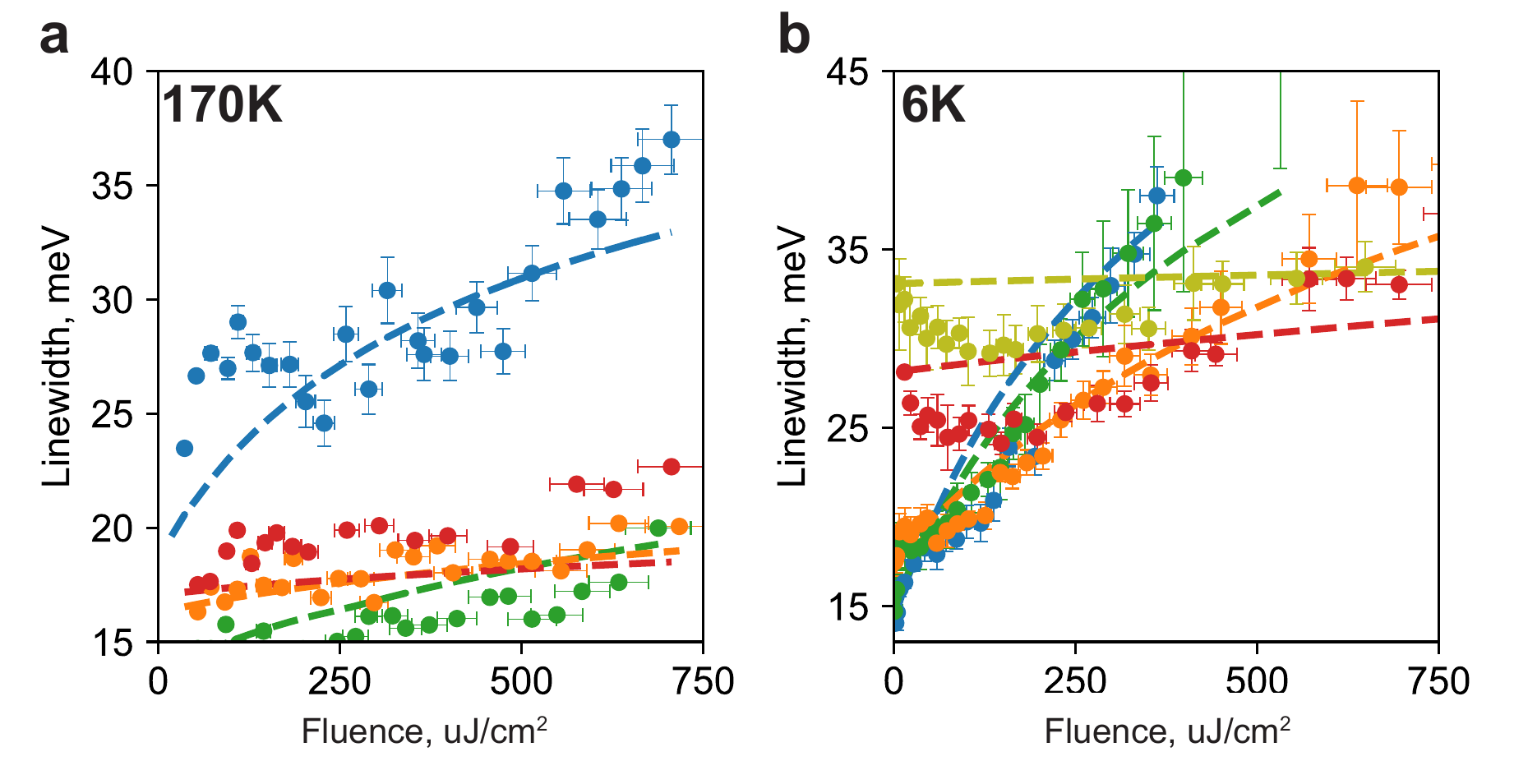}}
\caption{Extracted polariton mode linewidth under femtosecond laser excitation as a function of incident pump fluence at 170K (a) and 6K (b). Corresponding theoretical calculations are shown with dashed lines. Colors correspond to the detunings denoted in Fig \blue{3e,f}. Horizontal error bars are the root mean square error of the pump laser fluence originating from the laser instability. Vertical error bars correspond to the standard deviation error of Fano lineshape fitting.
}
\label{fig:S blueshift}
\end{figure}

\section{Influence of thermal effects on the blueshift measurements}
\begin{figure}[t!]
\centering
\center{\includegraphics[width=0.8\linewidth]{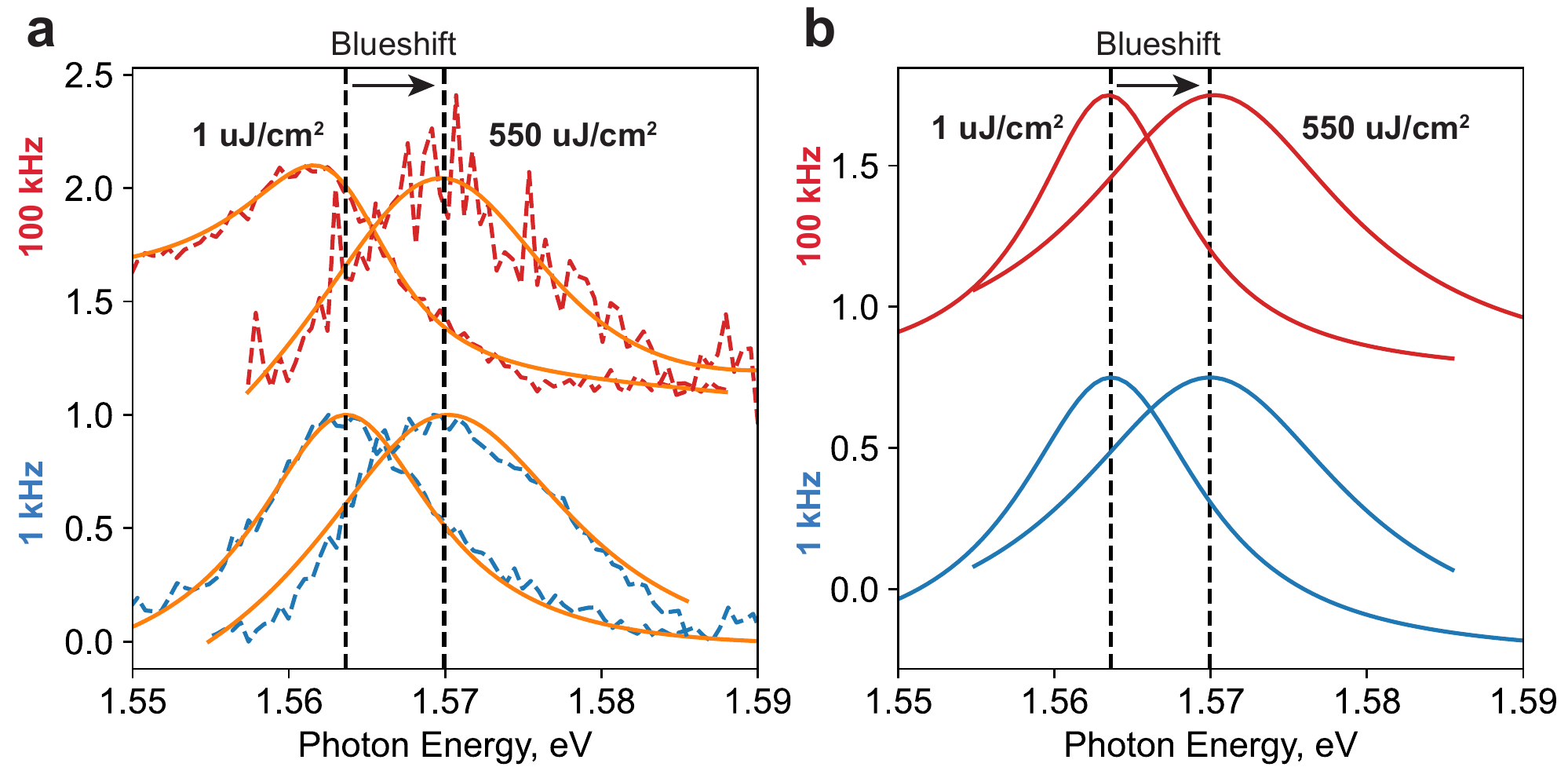}}
\caption{Reflectance spectra of polariton mode measured at 170K under resonant pump for detuning corresponding to $|X|^2 = 0.25$. (a) Dashed lines represent extracted polariton reflectance spectra under resonant pump at 1 and 550 uJ/cm$^2$. Solid orange lines show the fitted Fano Resonance functions. Red color corresponds to 100 kHz of laser repetition rate, blue color corresponds to 1 kHz. (b) Lorenzian functions of polariton resonances with the parameters obtained from fitting in panel (a)}
\label{fig:SrrCheck}
\end{figure}

In order to confirm the absence of the thermal effects on the measured nonlinearities, which measure the optical nonlinear blueshifts at 170K for the the detuning of  $|X|^2 = 0.25$ with repetition rates of 100 and 1 kHz. The 100 kHz repetition rate was used in the main experiment, while 1 kHz was used to confirm that the optical response of the sample remains the same, which would exclude any temperature effects. Using the method described in the main text, we measure polariton peak in the reflectance spectra under the resonance pump at incident fluences of 1 uJ/cm$^2$ and 550 uJ/cm$^2$ and fit the obtained spectra by the Fano resonance function (Fig~\ref{fig:SrrCheck}a). To see it more clearly, we plot Lorentzian peak functions with the extracted parameters to get rid of the asymmetry in the peak shapes (Fig \ref{fig:SrrCheck}b). In these two experiments, the asymmetry parameter slightly differs due to the influence of an additional neutral density filter used in the setup. The polariton mode peak positions and linewidths measured for two different incident fluences, perfectly coincide for 1 and 100 kHz repetition rates, which confirms the absence of the thermal effects (such as sample heating) in our experiment.

\section{Theory of exciton-polariton nonlinearity}

\subsection{Exciton-exciton Coulomb interaction in hydrogenic picture}

Given by equal exciton center of mass momentum and parallel spin orientation of electrons and holes, the exciton-exciton interaction rate is the sum of four contributions \cite{ciuti1998role1}:

\begin{equation}
    V_{XX}(q) = V_d (q) + V_X (q) + V_e (q) + V_h (q),
\end{equation}
where the first term stands for direct interaction, and the remaining terms correspond to exciton, electron, and hole exchange, respectively. Here $q$ stands for exchange wave vector. 
The direct interaction plays an important role for spatially indirect excitons possessing dipole moment, while in typical conditions it is negligibly small. 
Typically the exchange  wavevector dependence of  interaction rate is neglected, and the interaction is well approximated as

\begin{equation}
V_{XX} = V_{XX}(0) \approx  2 V_e (0).
\end{equation}
The electron exchange interaction is defined as \cite{ciuti1998role1,glazov2009polariton}

\begin{align}
    V_e (0) = \frac{1}{L^{2d}} & 
\int d^d\mathbf{r}_e d^d \mathbf{r}_h d^d \mathbf{r}_{e'} d^d \mathbf{r}_{h'} \Psi^*(|\mathbf{r}_e-\mathbf{r}_h|) \Psi^*(|\mathbf{r}_{e'}-\mathbf{r}_{h'}|) 
\notag \\
& \left[ 
-V(\mathbf{r}_e - \mathbf{r}_e')
-V(\mathbf{r}_h - \mathbf{r}_h')
+V(\mathbf{r}_e - \mathbf{r}_h')
+V(\mathbf{r}_h - \mathbf{r}_e')
\right]
\Psi(|\mathbf{r}_{e'}-\mathbf{r}_h|) \Psi(|\mathbf{r}_e-\mathbf{r}_{h'}|) ,
\label{eq:Ve}
\end{align}
where $V (r) = e^2/{ (4 \pi \varepsilon_0 \varepsilon r)}$ stand for Coulomb interaction, $\psi$ is the wavefunction of exciton internal dynamics, $L^d$ is the normalization area [volume], and $d=2,3$ for 2D and 3D, respectively.
The direct calculation results in

\begin{align}
    V_e^{2D} & \approx 6 E_b^{2D} (a_B^{2D})^2 ,
    \label{eq:Ve2D}
\end{align}
\begin{align}    
    V_e^{3D} & \approx 27 E_b^{3D} (a_B^{3D})^3 .
    \label{eq:Ve3D}
\end{align}
Here the Bohr radius and binding energy read as

\begin{align}
    E_b^{3D} & = \frac{1}{(4\pi\varepsilon_0 \varepsilon)^2} \frac{\mu e^4 }{2 \hbar^2}, \\
    a_B^{3D} & = 4\pi\varepsilon_0 \varepsilon \frac{\hbar^2}{\mu e^2} ,
\end{align}
and $E_b^{2D} = 4 E_b^{3D}$, $a_B^{2D} = a_B^{3D}/2$.

The three-body Coulomb correlation term can be presented as \cite{combescot2008many1}:

\begin{align}
    V_{XX2} & = 
    \frac{1}{L^{4d}}
    \left(  \sum_{k,k^\prime} 2V_{\vec{k}-\vec{k}^\prime} |\psi_k|^4 \left( |\psi_{k^\prime}|^2 - \psi_{k^\prime} \psi_k^* \right) 
    - \frac{1}{L^{d}} \sum_q |\psi_q|^4 \sum_{k,k^\prime} 4 V_{\vec{k}-\vec{k}^\prime} |\psi_k|^2 \left( |\psi_{k^\prime}|^2 - \psi_{k^\prime} \psi_k^* \right) \right) ,
    \label{eq:U2}
\end{align}
where $V_k$ and $\psi_k$ are the Fourier images of the interaction potential and the exciton wavefunction, respectively.
The resulting expressions read as

\begin{align}
    V_{XX2}^{2D} & \approx - 59 E_b^{2D} (a_B^{2D})^4 , 
    \label{eq:U22D}
\end{align}
\begin{align}
    V_{XX2}^{3D} & \approx - 266 E_b^{3D} (a_B^{3D})^6 . 
    \label{eq:U23D}
\end{align}
In the expressions \eqref{eq:Ve2D}, \eqref{eq:Ve3D}, \eqref{eq:U22D}, \eqref{eq:U23D} we omitted the normalization factors.

\subsection{Exciton-polaron nonlinearity rates}

For the case of exciton-polaron we evaluate the nonlinearity rates by means of the Eqs. \eqref{eq:Ve}, \eqref{eq:U2}. 
Here the Coulomb interaction potential is given by the Eq. (7) of the main text, which in the inverse space reads

\begin{equation}
    V_k = \frac{e^2}{4\pi\varepsilon_0} \left[
    \frac{4 \pi }{\epsilon_s q^2 }
    + \frac{4 \pi }{\epsilon_p} \left( \frac{m_h}{\Delta m} 
    \frac{1}{\frac{1}{l_h^2}+q^2}
    -\frac{m_e}{\Delta m} 
    \frac{1}{\frac{1}{l_e^2}+q^2}
    \right) \right] .
\end{equation}
The exciton wave function is found in the form of conventional (3D) hydrogen-like shape, in the inverse space reading as 

\begin{equation}
    \psi_k =  \frac{8 \sqrt{\pi a_B^3} }
    {\left[1+ (a_B k)^2 \right]^2 }.
\end{equation}
The resulting values for 170K tetragonal phase exciton are  
$V_{e}^{T} = 0.0157$ $\mu$eV$\cdot \mu$m$^3$, 
$V_{XX2}^{T} = -2.5 \cdot 10^{-9}$ $\mu$eV$\cdot \mu$m$^6$.
Taking into account the values of binding energy $E_b^T = 19$ meV, and the Bohr radius $a_B^T = 2.36$ nm, we can write

\begin{align}
    V_{e}^{T} = 62.86 E_b^{T} (a_B^{T})^3
\end{align}
\begin{align}
    V_{XX2}^{T} = -761.57 E_b^{T} (a_B^{T})^6.
\end{align}

The expressions for saturation rates read as \cite{emmanuele2020highly1}

\begin{equation}
    s = 2\frac{\sum\limits_{k} |\psi_k|^2 \psi_k }{\sum\limits_{q} \psi_q} ,
\end{equation}
\begin{equation}
    s_2 = 2\frac{\sum\limits_{k} |\psi_k|^2 \psi_k \sum\limits_{k^\prime} |\psi_{k^\prime}|^4
    - \sum\limits_{k^{\prime \prime}} | \psi_{ k^{\prime \prime}} |^4 \psi_{k^{\prime \prime}} }
    {\sum\limits_{q} \psi_q} .
\end{equation}

\subsection{Blueshift comparison}

\begin{figure*}
\centering
\center{\includegraphics[width=0.4\linewidth]{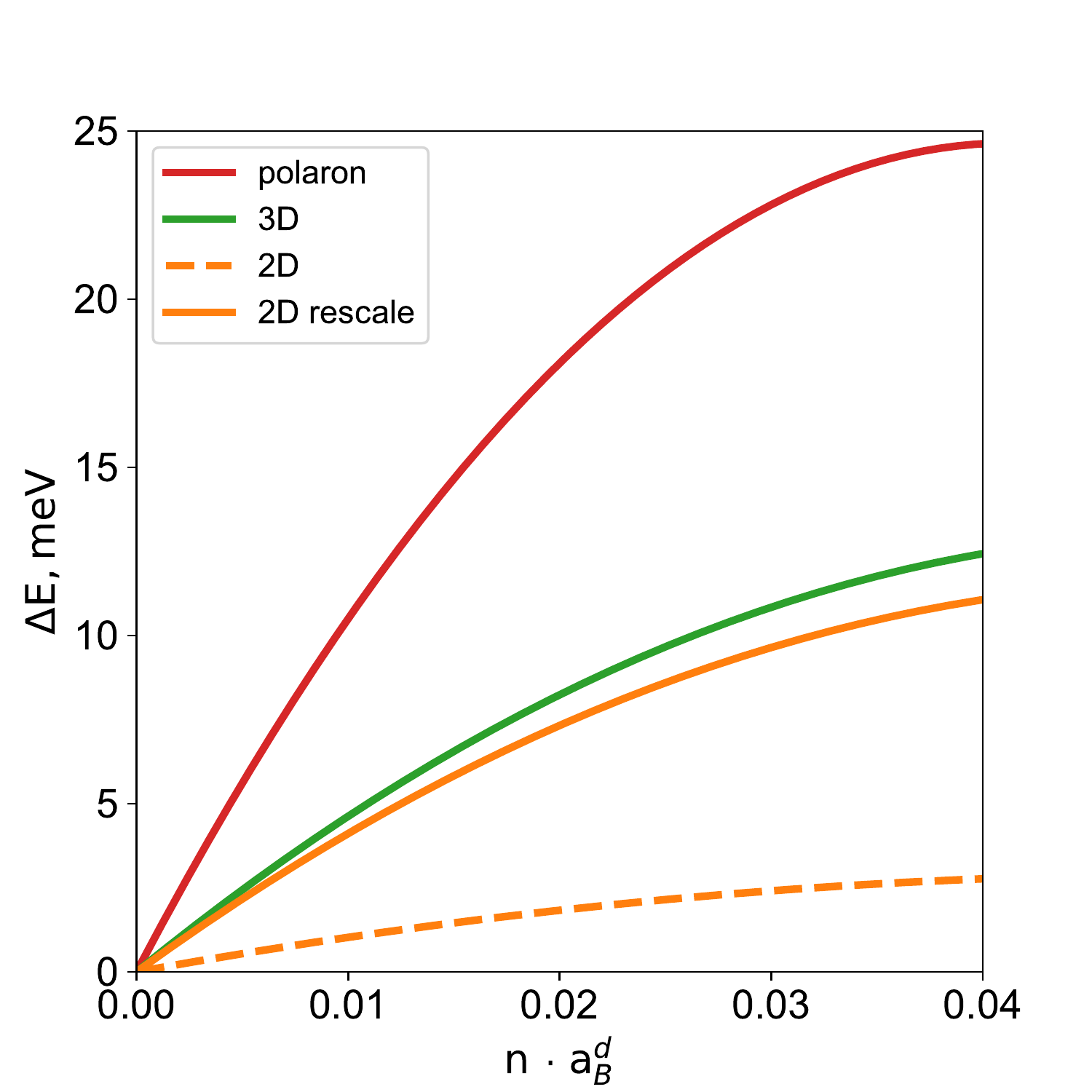}}
\caption{
Exciton spectrum blueshift versus the density in systems of different dimensionality. Here the exciton-polaron parameters $E_b = 19$ meV, $a_B = 2.36$ nm are used for the 2D and 3D models as well. The orange solid curve corresponds to rescaled 2D exciton (the binding energy and Bohr radius are taken as $4E_b$, $a_B/2$, respectively)}
\label{fig:blueshift}
\end{figure*}

In order to highlight the blueshift enhancement due to the polaron effects, we plot the density dependence of blueshift for various structures in Fig. \ref{fig:blueshift}. Here we choose the parameters of exciton-polaron in tetragonal phase, and compare with (2D) and (3D) hydrogen-like excitons having the same parameters. The results clearly demonstrate the strong enhancement of blueshift in the case of polaron renormalization of bulk excitons.

\section{The phonon energy dependence of exciton Coulomb nonlinearity}

\begin{figure*}
\centering
\center{\includegraphics[width=0.7\linewidth]{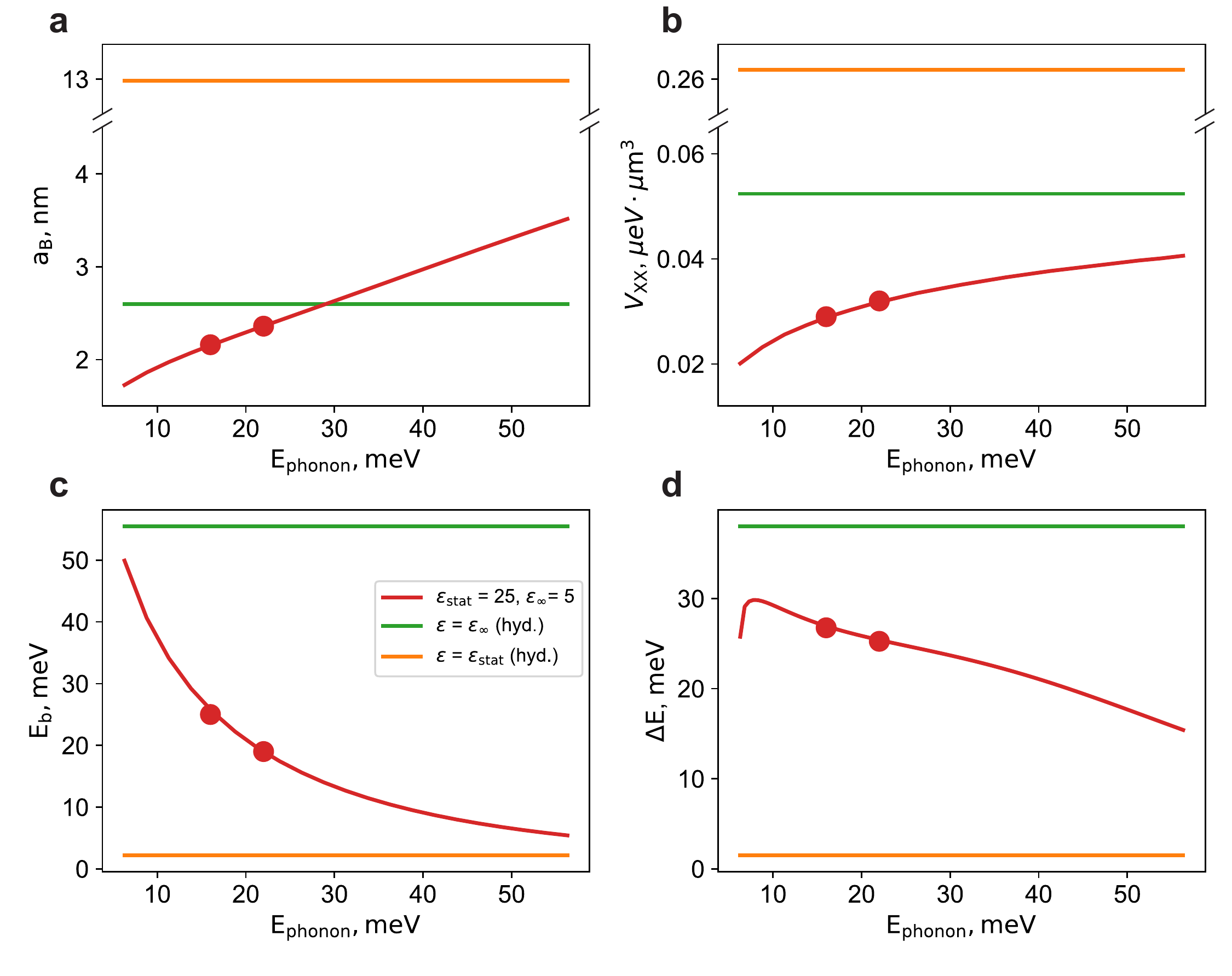}}
\caption{The exciton Bohr radius (a),  XX-interaction rate (b), binding energy (c), and effective blueshift at Mott transition density (d)  versus the phonon energy. 
The orange and green curves correspond to pure 3D hydrogen-type limits when the dielectric screening  $\epsilon = \epsilon_s$ and  $\epsilon = \epsilon_\infty$, respectively. The red dots correspond to values used to fit the experimental data.
Despite the reduction of Coulomb nonlinearity rate (panel b), the maximal possible blueshift is higher due to the increase of the maximum of particle density.
}
\label{figSexc}
\end{figure*}

The phonon energy dependence of Bohr radius and exciton binding energy is shown in Fig.~\ref{figSexc} a, c. 
At the limits $E_{ph} \rightarrow 0 [\infty]$ the exciton binding energy approaches 3D hydrogen model with dielectric permittivity $\varepsilon \rightarrow \varepsilon_{\infty[s]}$, respectively. 
However, the corresponding Bohr radius tends even below the hydrogen limit with low permittivity, which is due to the violation of Rydberg scaling. 
Because of the reduction of interaction cross-section, this results in the decrease of exciton-exciton Coulomb interaction  (see. Fig.~\ref{figSexc} b).
Despite this circumstance, the maximal possible blueshift is larger (see. Fig.~\ref{figSexc} d) due to the enhanced maximal density of particles.

\section{The temperature dependence of exciton Coulomb nonlinearity and light-matter coupling strength}

\begin{figure*}
\centering
\center{\includegraphics[width=0.7\linewidth]{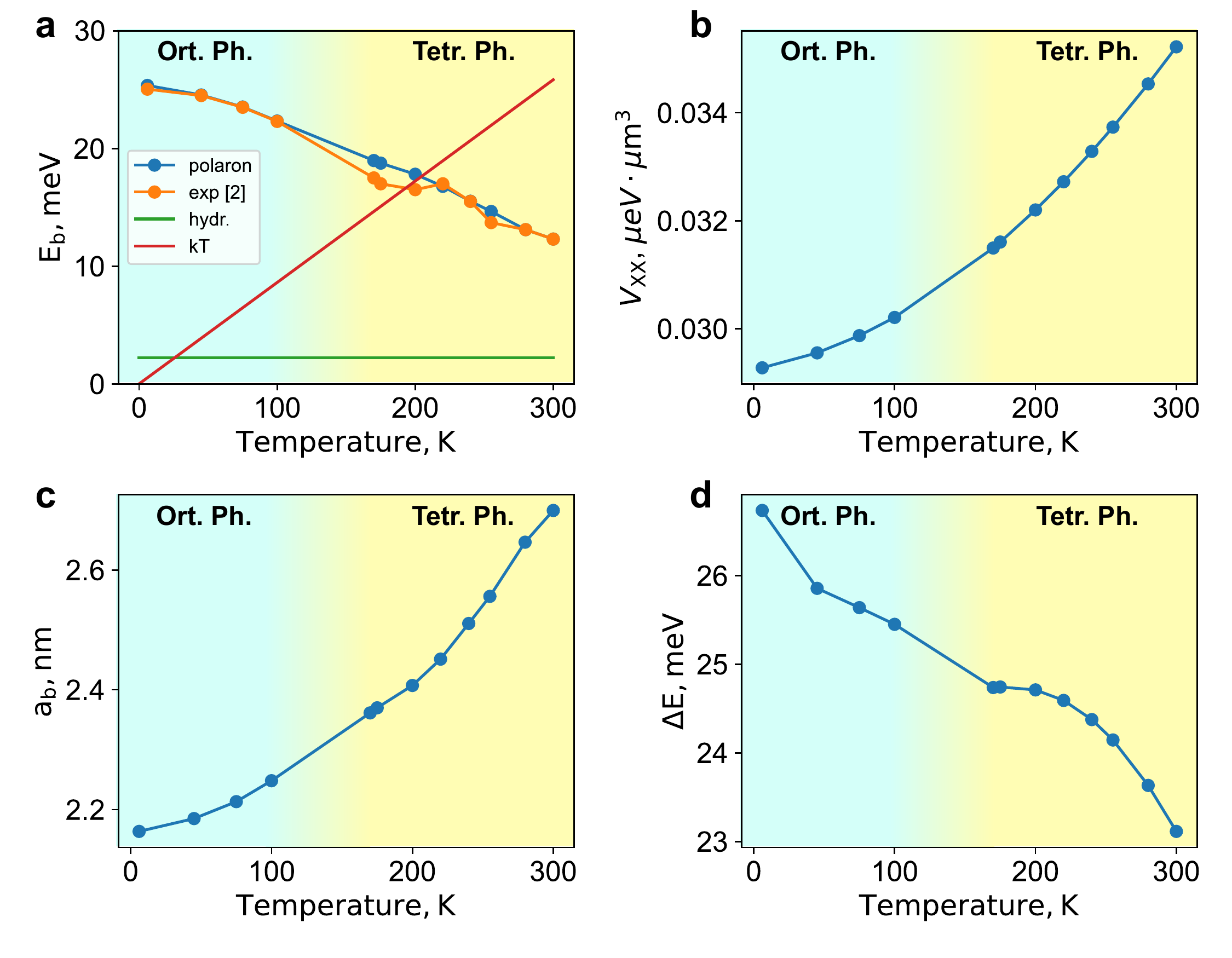}}
\caption{The temperature dependence of exciton binding energy (a), XX-interaction rate (b), Bohr radius (c), and maximal blueshift (d).
In panel (a) the blue curve is the exciton binding energy calculated within the exciton-polaron model, the orange curve corresponds to experimental data extracted from Ref. \cite{soufiani2015polaronic1}, the green curve is the 3D hydrogen model with $\varepsilon= \varepsilon_{s}$, and the red line is the thermal cut-off energy $k_B T$, with $k_B$ denoting the Boltzmann constant. 
Yellow and blue regions correspond to tetragonal and orthorhombic crystal phases, respectively.}

\label{figST}
\end{figure*}

The temperature dependence of excitonic and electronic properties of MAPbI$_3$ is studied in detail in Ref. \cite{soufiani2015polaronic1}. 
It is demonstrated, that the phase transition at around 160 K primarily manifests itself in the abrupt change of the bandgap. On the contrary, the excitonic binding energy varies rather smoothly, from $\sim25$ meV at 6 K to $\sim 12$ meV at room temperature. 
In Fig.~\ref{figST}a we present the extracted data of temperature dependence of exciton binding energy and the corresponding theoretical fit within the exciton-polaron treatment.
We highlight here, that the hydrogen-like exciton with binding energy $E_b = 2.2$ meV would be thermally unstable at 170 K due to the condition $E_b \ll k_B T$.

Exploiting the obtained exciton wave functions, we further calculate the exciton Bohr radius, nonlinearity rates associated with Coulomb correlations between the excitons, and the respective maximal blueshift. 
The results of calculation  are presented in Fig.~\ref{figST} b-d, demonstrating a moderate temperature dependence of maximal blueshift. 

\begin{figure*}
\centering
\center{\includegraphics[width=1\linewidth]{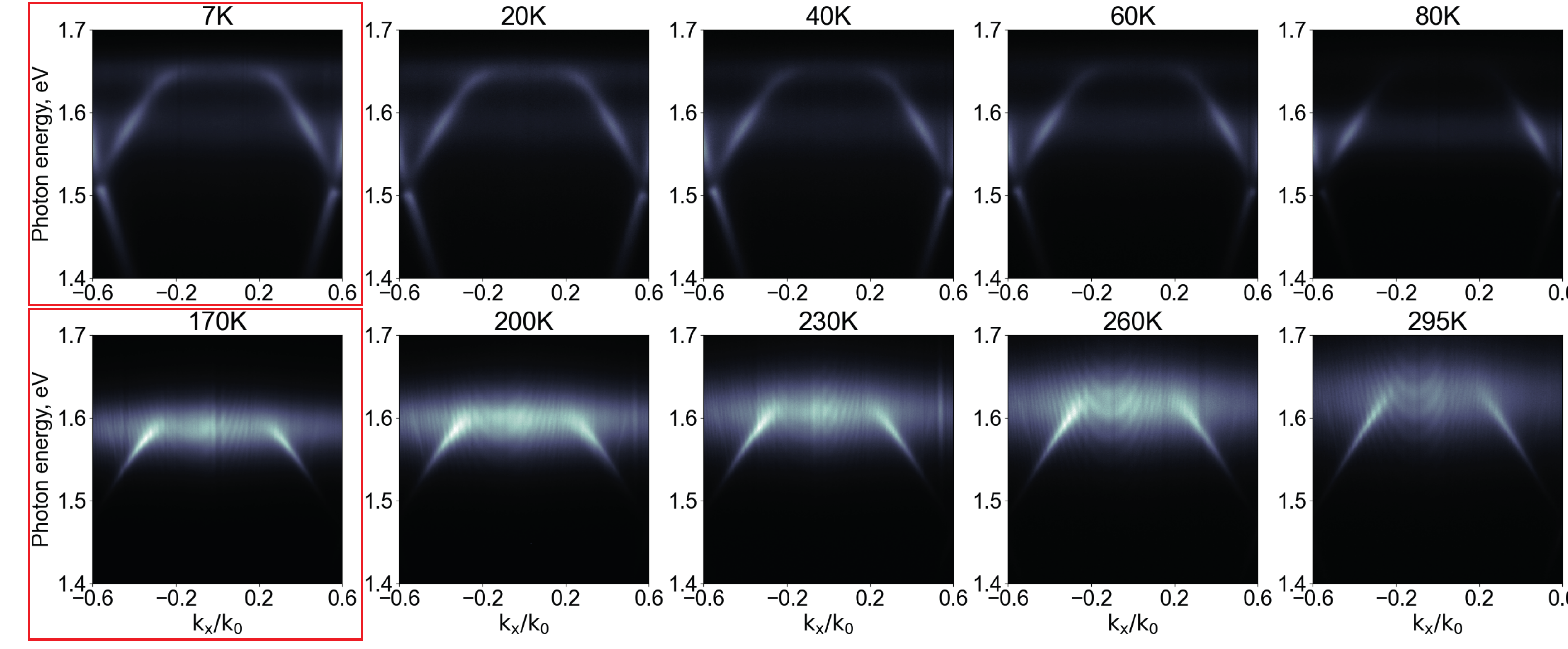}}
\caption{The evolution of angle-resolved PL spectra of the sample with varying the temperature. Measurements at 6K and 170K used in the main text are highlighted with red squares.}
\label{figBFPplT}
\end{figure*}

\begin{figure*}
\centering
\center{\includegraphics[width=0.9\linewidth]{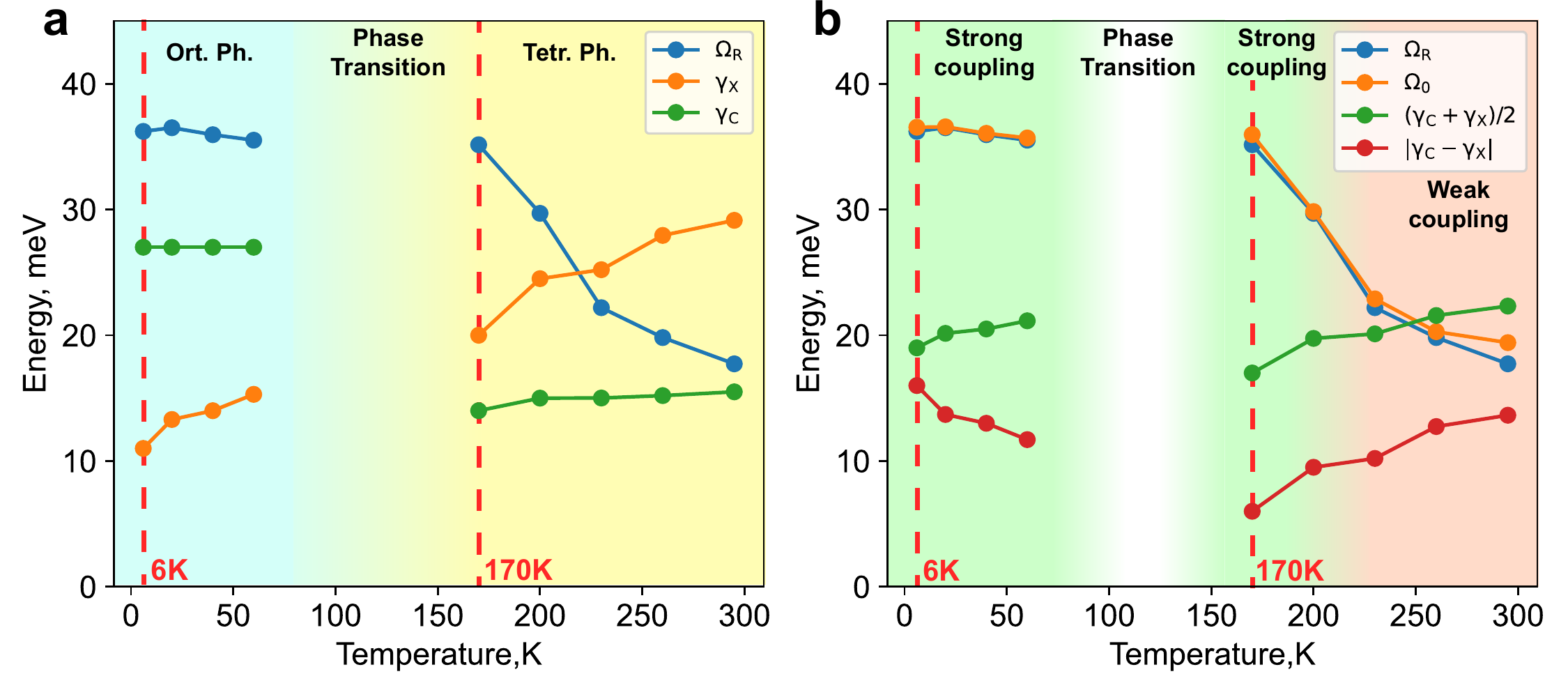}}
\caption{(a) Temperature dependencies of Rabi splitting, linewidths of uncoupled photon cavity mode and uncoupled exciton resonance.  (b) Comparison of the Rabi splitting energy and light-matter strength with half-sum and difference of cavity photon and exciton linewidths. In panel (a) yellow and blue regions correspond to tetragonal and orthorhombic crystal phases respectively. In panel (b) the red region corresponds to the weak light-matter coupling regime, green region corresponds to the regime of strong light-matter coupling: $\Omega_0 > |\gamma_C - \gamma_{X}|$ and $\Omega_R > (\gamma_C + \gamma_{X})/2$}
\label{figSRabiT}
\end{figure*}

In the study, we measured angle-resolved PL spectra at different temperatures, shown in Fig \ref{figBFPplT}, from which we extract polariton branches and fit them by the two-coupled oscillator model, as it was described in the Methods section in the main text. 
The resulting Rabi splitting, linewidths of uncoupled cavity mode, and uncoupled exciton is shown in Fig \ref{figSRabiT}, a. 
To check the strong light-matter coupling regime we plot Rabi splitting, light-matter coupling strength, half-sum and full-difference of cavity photon and exciton linewidths in Fig \ref{figSRabiT}, b. 
In the red region, where half-sum of linewidths exceeds Rabi splitting we claim a weak light-matter coupling regime. 
The green region, where linewidths half-sum is lower than Rabi splitting, corresponds to a strong light-matter coupling regime. 
It should be noticed, that there is the phase transition, where it is impossible to clearly determine polariton branches and therefore we do not determine data.

\begin{figure*}
\centering
\center{\includegraphics[width=0.7\linewidth]{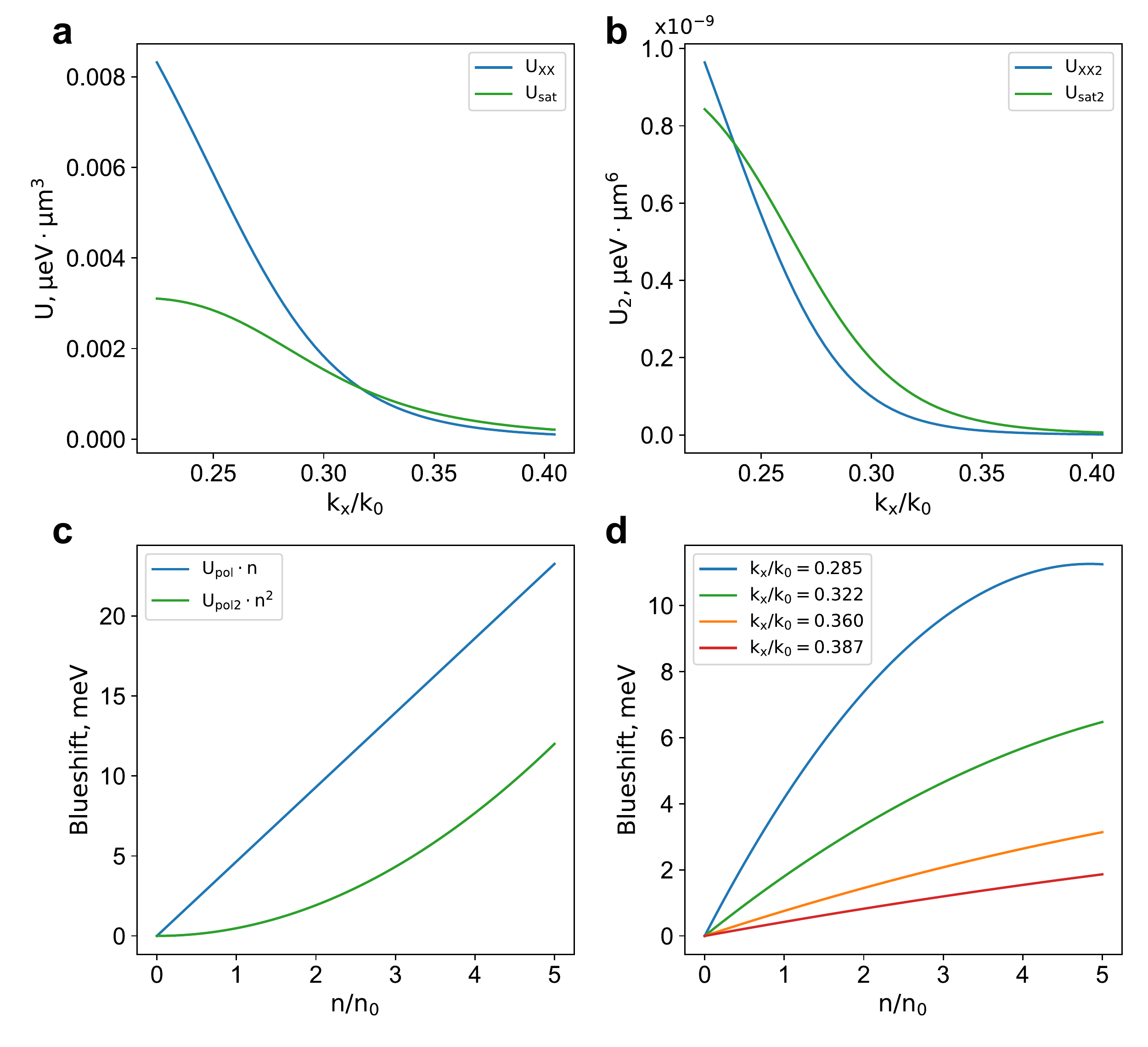}}
\caption{
Wavevector dependence of 
the first (a), and the second (b) order polariton nonlinearity associated with exciton-exciton interaction (XX), and the quench of Rabi splitting (sat).
(c) Density dependent energy shifts of polariton energy associated with the first and the second order nonlinearity. Here $k_x=0.285k_0$, and $n_0 = 10^{18}$ cm$^{-3}$.
(d) Density dependent blueshift of polariton energy at different values of wavevector.  
}
\label{figSnon}
\end{figure*}

\section{Polariton nonlinearity rates}

The blueshift of lower polariton branch with increased pump fluence is defined by Eqs. (6) and (7) of the main text. The contributions of nonlinearity rates associated with exciton-exciton interactions and the quench of Rabi splitting are

\begin{align}
    \label{eq:U1c}
    U_{XX} (k) & \approx 
    \frac{V_{XX}}{2} |X_k|^4 ,\\
    U_{\rm sat} (k) & = \frac{\Omega_0 }{2} s |X_k|^2 (X_k^* C_k + X_k C_k^*),
\end{align}
\begin{align}
    \label{eq:U2c}
    U_{XX2} (k) & \approx 
    |V_{XX2}| |X_k|^6 , \\
    U_{\rm sat 2} (k) &= \frac{\Omega_0 }{2} |s_2| |X_k|^4 (X_k^* C_k + X_k C_k^*).
\end{align}
In the Fig.~\ref{figSnon} a, b the comparison of two effects for the first and second order of polariton nonlinearity is presented. 
While for the first order the Coulomb contribution is larger at moderate values of wavevector, the second order term is fully dominated by the quench of Rabi splitting. 
We mention that within out treatment the wavevector dependence stems from the Hopfield coefficients solely. 
In the Fig.~\ref{figSnon} c the modulus of energy shifts arising from the first and second order nonlinear terms versus the particle density is presented. 
The contribution of the second order term results in the deviation from linear scaling at elevated particle densities, as illustrated in Fig.~\ref{figSnon} d.

\end{document}